\documentclass[authoryear]{elsarticle}

\usepackage{amsmath}
\usepackage{bm}
\usepackage{tikz}
\usepackage{pgfplots}
\usepackage{color, colortbl}
\usepackage{graphicx}

\graphicspath{{pictures/}}

\newproof{pf}{Proof}

\journal{arXiv.org}

\begin{document}

\begin{frontmatter}

\title{State change modal method for numerical simulation of dynamic processes in a nuclear reactor}

\author[ki]{Alexander V. Avvakumov}
\ead{Avvakumov2009@rambler.ru}

\author[nsi]{Valery~F.~Strizhov}
\ead{vfs@ibrae.ac.ru}

\author[nsi,univ]{Petr N. Vabishchevich\corref{cor}}
\ead{vabishchevich@gmail.com}

\author[univ]{Alexander O. Vasilev}
\ead{haska87@gmail.com}

\address[ki]{National Research Center \emph{Kurchatov Institute},  1, Sq. Academician Kurchatov, Moscow, Russia}
\address[nsi]{Nuclear Safety Institute, Russian Academy of Sciences, 52, B. Tulskaya, Moscow, Russia}
\address[univ]{North-Eastern Federal University, 58, Belinskogo, Yakutsk, Russia}

\cortext[cor]{Corresponding author}

\begin{abstract}
Modeling of dynamic processes in nuclear reactors is carried out, mainly, on the basis of the multigroup diffusion approximation for the neutron flux. The basic model includes a multidimensional set of coupled parabolic equations and ordinary differential equations. Dynamic processes are modeled by a successive change of the reactor states, which are characterized by given coefficients of the equations. In the modal method, the approximate solution is represented as an expansion on the first eigenfunctions of some spectral problem. The numerical-analytical method is based on the use of dominant time-eigenvalues of a multigroup diffusion model taking into account delayed neutrons. Numerical simulations of the dynamic process were performed in the framework of the two-group approximation for the VVER-1000 reactor test model. The last is characterized by the fact that some eigenvalues are complex.
\end{abstract}

\begin{keyword}
Neutron diffusion equations \sep  multi-group approximation \sep space-time kinetic 
\sep spectral problem \sep modal method.

\end{keyword}

\end{frontmatter}

\section{Introduction} 

Among the basic physical processes occurring in a nuclear reactor  \citep{duderstadt1976nuclear}, 
the focus is on neutron transport. To describe that process, a rather complex integro-differential equation is involved, in which the neutron flux distribution depends on time, energy, spatial and angular variables  \citep{hetrick1971dynamics,stacey}. 
In practical calculations of nuclear reactors, as a rule,  simpler systems of equations in the multigroup diffusion  approximation  are used  \citep{marchuk1986numerical,lewis1993computational,sutton1996diffusion,cho2005fundamentals}.
Currently diffusion models are derived and applied using sophisticated homogenization methodologies (see \cite{sanchez2009assembly,dugan2016cross}) 
which define parameters of the multigroup diffusion equations that enable one to take into account transport effects. 

The computational model is a system of coupled second-order parabolic equations, which is supplemented by a system of ordinary differential equations to take into account delayed neutrons. Engineering neutronics codes designed to simulate neutron transport in the few-group diffusion approximation are based, mainly, on finite-difference spatial approximations. To improve the calculation accuracy, nodal methods are widely used (see, for example, \cite{smith1979analytic,lawrence1986progress}). 
Applying these methods, a rather coarse grid can be used: several points per assembly in plane and several dozen layers in height. Nodal methods are based on the representation of the neutron flux within the computational element as a small degree polynomial or a set of functions along one of the coordinates (or in plane). In some cases, nodal methods can be related  \citep{grossman2007nodal} with special variants of finite-element approximation. It should be noted that to increase the finite-element accuracy of approximate solution of boundary value problems, it is more appropriate to use standard procedures with condensed  computational grids finite elements of higher degree. Such technology is used in  \cite{vidal2014solution,avvakumov2017spectral} 
when considering spectral problems for the multigroup diffusion equations.
Modern computational algorithms for solving spectral problems in neutron theory are discussed by \cite{gill2011newton,willert2014comparison}. 

When modeling the dynamics of neutron-physical processes, standard methods for the approximate solution of non-stationary problems are used  \citep{sutton1996diffusion,cho2005fundamentals,stacey}. The most attention is paid to two-layer schemes with weights ($\theta$-method) \citep{Ascher2008,LeVeque2007,HundsdorferVerwer2003},
also Runge-Kutta and Rosenbroek schemes \citep{Butcher2008,HairerWanner2010} are used. To characterize the reactor dynamics, the spectral parameter  $\alpha$ is used. It is defined as the main eigenvalue of the time-eigenvalue  ($\alpha$-eigenvalue) problem, which is associated with the nonstationary neutron diffusion equations 
\citep{Bell1970,modak2007scheme,verdu20103d}.
By analogy with the usual problems of heat conduction (see, for example,  \cite{luikov2012analytical,samarskii1996computational}) we can discriminate the regular regime of the reactor. For large times, the neutron flux behavior has an asymptotic character when one can speak of the space-time factorization of a solution whose amplitude is $\exp(-\alpha t)$, the form-function is the spectral problem eigenfunction. To model the reactor regular regime, it is necessary to focus on the use of completely implicit schemes, while the Crank-Nicholson scheme is unsuitable  \citep{nd-mm}.

Real three-dimensional dynamic neutronic calculations require the use of large computational grids for large characteristic times which in turn determine the  use of modern multiprocessor computer systems. Parallel computational algorithms are based on a sequence of solving more simple problems for individual processes. Advance is achieved through the use of decoupling technology: splitting by physical processes 
\citep{Vabishchevich2014}), decomposition of the computational domain into subdomains  \citep{ToselliWidlund2005},
iterative methods for solving systems of algebraic equations  \citep{Saad2003}. 
In the case of spectral problems for the neutron diffusion, domain decomposition methods are used, for example, in  \cite{guerin2010domain}. 
Features of solving non-stationary problems on parallel computers are taken into account by constructing special iterative methods such as the parareal in time algorithm  \citep{maday2005parareal}. 
This approach was implemented \cite{baudron2014parareal} for numerical calculation of transient multigroup neutron kinetics equations involving a time delayed contributions.

In the theory and practice of neutronics calculations, fast methods to obtain approximate solutions are given great attention. In this connection, a class of methods for modeling the nonstationary group neutron diffusion should be noted. It is associated with the multiplicative representation of the space-time factorization methods and the quasistatic method  \citep{dodds1976accuracy,chou1990three,goluoglu2001time,dulla2008quasi,dahmani20013d}.
In this case, an approximate solution is represented as a product of two functions: one of which depends on the time and is related to the amplitude, the second (shape function) describes the spatial distribution. The shape function is often associated with the fundamental eigenfunction of certain eigenvalue problems for neutron diffusion equations. 

When using the quasistatic method, the problem is significantly simplified, thus, it is doubtful to obtain good accuracy for an approximate solution, in particular, in dynamic regimes with a complex redistribution of the neutron flux. For this reason, the more general approach of the modal method has been successfully developed  \citep{stacey1967modal,stacey1969space,sutton1996diffusion}.
In this case, the solution is represented in the form of a sum of several dominant eigenvalues with time-dependent coefficients.

To characterize the reactor dynamic processes some spectral problems  \citep{Bell1970,hetrick1971dynamics,stewart1976spectral,stacey} were considered. The processes occurring in a nuclear reactor are essentially non-stationary. 
The stationary state of neutron flux is characterised by local balancing of neutron absorption and generation
and is usually described by solution of a spectral problem (Lambda modes problem, $\lambda$-eigenvalue problem).
The fundamental eigenvalue (maximal eigenvalue) is called $k$-effective of the reactor core.
The nodal method based on the use of the  $\lambda$-eigenvalue problem is discussed, for example,  \cite{verdu1998modal,miro2002nodal,gonzalez2009high}. In particular, it should be paid attention to issues related with the calculation of the related system of equations for the nonstationary expansion coefficients. 

Nonstationary processes can naturally be described on the basis of the approximate solution expansion in time-eigenvalue of  $\alpha$-eigenvalue problem \citep{ginestar2002transient,verdu20103d,verdu2014modal}.
In a simpler model without delayed neutrons, modal methods were used in  \cite{modak2007scheme}.
The principal point is connected with the fact that when using this approach, we deal with an very big system of equations for the coefficients. It should also be noted that the eigenvalues are complex for both   $\lambda$- and 
$\alpha$-eigenvalue problem. To set the initial state, this leads to the need to solve the appropriate adjoint spectral problems.

In this paper, we formulate a general strategy for the approximate solution of nonstationary problems of neutron transport in nuclear reactors, which is oriented to fast real-time calculations using the State Change Modal (SCM) method. The dynamic behavior of a nuclear reactor is considered as a sequence of its states characterized by its set of constant coefficients of the multigroup diffusion equations. It is considered that the transition from one state to another occurs instantaneously. For a separate state, the neutron flux is calculated using the modal method to represent the problem’s solution in the form of expansion in the dominant eigenfunctions of the  $\alpha$-eigenvalue problem
taking into account delayed neutrons (the full Alpha method). Real-time calculations are provided by the fact that the desired set of eigenvalues and eigenfunctions of the reactor deterministic state are calculated in advance.

The paper is organized as follows. The dynamic model of a nuclear reactor based on the multigroup diffusion equations is given in Section 2. The general strategy of numerical modeling of nonstationary processes based on the SCM method is described in Section 3. The key computational aspects of the SCM technology are discussed in Section 4. Two-dimensional test problem for VVER-1000 reactor is discussed in Section 5. Modeling of the dynamic process, which corresponds to two reactor states (supercritical and subcritical): a transition from the regular regime of the critical state to the subcritical state of the reactor. The results of the work are summarized in Section 6.

\section{Problem statement}

The neutron flux is modelled in multigroup diffusion approximation. The neutron dynamics is considered in the bounded convex two-dimensional or three-dimensional area  $\Omega$ ($\bm x = \{x_1, ..., x_d\} \in \Omega, \ d = 2,3$) with boundary $\partial \Omega$. The neutron transport is described by the system of equations:
\begin{equation}\label{1}
\begin{split}
 \frac{1}{v_g} \frac{\partial \phi_g}{\partial t} - & \nabla \cdot D_g \nabla \phi_g + \Sigma_{rg} \phi_g 
 - \sum_{g\neq g'=1}^{G} \Sigma_{s,g'\rightarrow g} \phi_{g'} \\
 =  & \ (1-\beta) \chi_g \sum_{g'=1}^{G} \nu \Sigma_{fg'} \phi_{g'} + \widetilde{\chi}_g \sum_{m=1}^{M} \lambda_m c_m , \quad 
 g = 1,2, ..., G .
\end{split}
\end{equation} 
Here $\phi_g(\bm x,t)$ --- neutron flux of $g$ group at point $\bm x$ and time $t$,
$G$ --- number of energy groups,
$v_g$ --- effective velocity of neutrons in the group $g$,
$D_g(\bm x)$ --- diffusion coefficient, $\Sigma_{rg}(\bm x,t)$ --- removal cross-section,
$\Sigma_{s,g'\rightarrow g}(\bm x,t)$ --- scattering cross-section from group $g'$ to group $g$,
$\beta$ --- effective fraction of delayed neutrons, $\chi_g$, $\widetilde{\chi}_g$  --- spectra of prompt and delayed neutrons, 
$\nu\Sigma_{fg}(\bm x,t)$ --- generation cross-section of group $g$,
$c_m$ --- density of sources of delayed neutrons of $m$-type,  $\lambda_m$ --- decay constant of sources of delayed neutrons,
$M$ --- number of types of delayed neutrons.
The density of sources of delayed neutrons is described by the equations:
\begin{equation}\label{2}
 \frac{\partial c_m}{\partial t} + \lambda_m c_m = \beta_m \sum_{g=1}^{G} \nu \Sigma_{fg} \phi_g ,
 \quad m = 1,2, ..., M , 
\end{equation} 
where $\beta_m$ is a fraction of delayed neutrons of $m$-type, and
\[
 \beta = \sum_{m=1}^{M} \beta_m .
\] 
System of equations (\ref{1}), (\ref{2}) is supplemented with corresponding initial and boundary conditions.

The albedo-type conditions are set at the boundary $\partial \Omega$ of the area $\Omega$:
\begin{equation}\label{3}
 D_g\frac{\partial \phi_g}{\partial n} + \gamma_g \phi_g = 0, \quad 
 \quad g = 1,2, ..., G ,
\end{equation}
where $n$ --- outer normal to the boundary $\partial \Omega$.
Initial state is defined in the following manner:
\begin{equation}\label{4}
 \phi_g(\bm x,0) = \phi_g^0(\bm x), 
 \quad c_m(\bm x,0) = c_m^0(\bm x) . 
\end{equation} 

Let's write the boundary problem (\ref{1})--(\ref{4}) in operator form. The vectors 
 $\bm \phi = \{\phi_1, \phi_2, ..., \phi_G\}$, $\bm c = \{c_1, c_2, ..., c_M\}$ 
and matrices are defined as follows:
\[
\begin{aligned}
 V & = (v_{g g'}), &
  \quad v_{g g'} & = \delta_{g g'} v_g^{-1}, \\
 D & = (d_{g g'}), &
 \quad d_{g g'} & = - \delta_{g g'} \nabla \cdot D_g \nabla, \\
 S & = (s_{g g'}), &
 \quad  s_{g g'} & =  \delta_{g g'} \Sigma_g - \Sigma_{s,g'\rightarrow g}, \\
 R & = (r_{g g'}), &
 \quad  r_{g g'} & = (1-\beta)\chi_g \nu \Sigma_{fg'}, \\
 B & = (b_{g m}), &
 \quad b_{g m} & = \widetilde{\chi}_g \lambda_m, \\
 \Lambda & = (\lambda_{m m'}), &
 \quad  \lambda_{m m'} & = \lambda_m \delta_{m m'}, \\
 Q & = (q_{mg}), &
 \quad  q_{mg} & = \beta_m \nu \Sigma_{fg}, \\
 g, g' & = 1,2, ..., G, &
 \quad m, m'  &= 1,2, ....,M,  
\end{aligned}
\]
where
\[
 \delta_{g g'} = \left \{ 
 \begin{matrix}
 1, & g = g', \\
 0, & g \neq  g',
 \end{matrix}
 \right . 
\] 
is the Kronecker symbol.
We shall use the set of vectors $\bm \phi$, whose components 
satisfy the boundary conditions (\ref{3}). 
Using the set definitions, the system of equations (\ref{1}), (\ref{2})  
can be written in the form of first-order equation of evolution:
\begin{equation}\label{5}
\begin{split}
V(t) \frac{d \bm \phi}{d t} + (D(t)+S(t)) \bm \phi &= R(t) \bm \phi + B(t)\bm c,
\\
\frac{d \bm c}{d t} + \Lambda(t)\bm c &= Q(t) \bm \phi. 
\end{split}
\end{equation}  
The Cauchy problem is formulated for equations (\ref{5})  when 
\begin{equation}\label{6}
 \bm \phi(0) = \bm \phi^0,
 \quad   \bm c(0) = \bm c^0,
\end{equation} 
where taken into account (\ref{4}) $\bm \phi^0 = \{ \phi_1^0,  \phi_2^0, ...,  \phi_G^0 \}$,  $\bm c^0 = \{c_1^0, c_2^0, ..., c_M^0\}$.

For an approximate solution of the Cauchy problem  (\ref{5}), (\ref{6}) 
two basic approaches are used. The first of these is connected with the use of standard two-layer or three-layer schemes, which are widely used in the numerical solution of parabolic problems  \citep{Samarskiibook}.

The operator matrices  $V, D$ are diagonal, and  $S$ is the lower triangular matrix. The essential binding of the equations is due only to the neutron generation operator $R$.
The problem of choosing design schemes amongst stable difference schemes, which is optimal for some additional criteria, is topical. In the theory of difference schemes, a class of asymptotically stable difference schemes is distinguished, which  \citep{samarskii1996computational} provide the correct behavior of the approximate solution for large times. In the theory of numerical methods for solving systems of ordinary differential equations  \citep{Butcher2008,Gear1971} the concept of  $L$-stable methods is introduced, in which, from several other positions, the asymptotic behavior of the approximate solution is also monitored for large times. A completely implicit scheme has better asymptotic properties than a symmetric scheme (the Crank-Nicholson scheme)  (see \cite{VabishchevichSM}), 
which is important in the study of the regular regime of a nuclear reactor \citep{nd-mm}.

The second class is the numerical-analytical methods for solving problem  (\ref{5}), (\ref{6}),
the most striking example of which are the modal methods noted above \citep{stacey1967modal,stacey1969space,sutton1996diffusion}.
They take into account the linear nature of the problem under consideration, the independence of the coefficients of the system of equations as a function of time. This allows us to construct an approximate solution by the method of separation of variables in the numerical determination of the dependence of the solution on the spatial variables.

\section{State change modal method} 

The nuclear reactor is always non-stationary. The limiting case of reaching a steady state (critical reactor) is observed only for certain coefficients of the equations system (\ref{5}). 
We will use the following simplified description of the dynamic processes in a nuclear reactor.

In selected time interval, the non-stationary neutron flux is determined by the nuclear reactor state. The state of the reactor is characterized by the constant coefficients of the system of multigroup diffusion equations (\ref{1}), (\ref{2}).

\begin{figure}[ht] 
  \begin{center}
\vspace{5mm} 
    \begin{tikzpicture}
      \filldraw [color=green!15] (0,0) rectangle +(9,1);
      \draw [dotted, line width=1, color=blue] (-1,0) -- (0,0);
      \draw [line width=1, color=blue] (0,0) -- (9,0);
      \draw [dotted, line width=1, color=blue] (9,0) -- (10,0);
      \draw [->, line width=1, color=blue] (10,0) -- (11,0);
      \filldraw [black] (3,0) circle (0.05);
      \filldraw [black] (6,0) circle (0.05);
      \draw  (1.5,0.5) node {state $s-1$};  
      \draw  (4.5,0.5) node {state $s$};  
      \draw  (7.5,0.5) node {state $s+1$};  
      \draw  (3.1,-0.4) node {$t_{s-1}$}; 
      \draw  (6.1,-0.4) node {$t_{s}$}; 
      \draw  (10.6,-0.4) node {$t$}; 
      \draw [->, line width=1, color=red] (3,1) -- (3,0);
      \draw [->, line width=1, color=red] (6,1) -- (6,0);
    \end{tikzpicture}
    \caption{State change scheme.} 
   \label{fig:1}
  \end{center}
\end{figure}
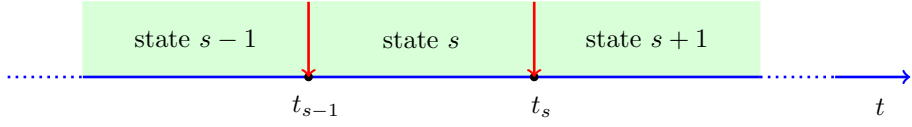

Dynamic processes in a nuclear reactor can be considered as a change of states  (see  Fig.\ref{fig:1}). 
At a certain time $t = t_s, \ s = 1,2, ...$ an instantaneous change of state occurs. The state
 $s$ is defined by the parameters in equations  (\ref{5}):
\[
 V(t) = V(t_s), \quad  D(t) = D(t_s), \quad  S(t) = S(t_s), \quad  R(t) = R(t_s), \quad  B(t) = B(t_s),
\] 
\[
 \Lambda(t) = \Lambda(t_s), \quad  Q(t) = Q(t_s),
 \quad t_{s-1} < t \leq t_s, \quad s = 1,2, ... .
\]

Simulation of the dynamic behavior of the reactor consists in solving the sequence of subtasks for the individual states of the reactor. The initial condition for the state $s$ (at $t = t_{s-1}$) is the final state of the reactor for the state $s-1$.

An approximate description of the non-stationary process at a separate stage is based on modal approximation. An approximate solution is sought in the form of decomposition in eigenfunctions of time and $\alpha$-eigenvalue problem. Finite-element approximation in space is used.

In a separate stage $s$ the following system of equations is considered
\begin{equation}\label{7}
\begin{split}
V(t_s) \frac{d \bm \phi}{d t} + (D(t_s)+S(t_s)) \bm \phi &= R(t_s) \bm \phi + B(t_s)\bm c,
\\
\frac{d \bm c}{d t} + \Lambda(t_s)\bm c &= Q(t_s) \bm \phi, 
\quad t_{s-1} < t \leq t_s,
\end{split}
\end{equation} 
supplemented by the corresponding initial conditions 
\begin{equation}\label{8}
 \bm \phi(t_{s-1}) = \bm \phi^s,
 \quad   \bm c(t_{s-1}) = \bm c^s .
\end{equation} 

Let's $\bm u = \{\bm \phi, \bm c\}$. Rewrite the system of equations (\ref{7}) as
\begin{equation}\label{9}
 \bm B \frac{d \bm u}{d t} + \bm A \bm u = 0,
 \quad t_{s-1} < t \leq t_s,
\end{equation} 
with constants
\[
 \bm A = 
 \begin{pmatrix}
 D(t_s)+S(t_s) - R(t_s) &  - B(t_s) \\
 - Q(t_s) & \Lambda(t_s) 
 \end{pmatrix} ,
 \quad  \bm B = 
 \begin{pmatrix}
 V(t_s) & 0 \\
 0 & I 
 \end{pmatrix} ,
\] 
where $I$ is the identity matrix. From (\ref{8}) one can obtain
\begin{equation}\label{10}
 \bm u(t_{s-1}) = \bm u^s .
\end{equation} 

After approximating over the space by the finite volume method or by the finite element method from (\ref{9}), (\ref{10}) 
we turn to the Cauchy problem for a linear system of ordinary differential equations with constant coefficients:
\begin{equation}\label{11}
 \bm B_h \frac{d \bm u_h}{d t} + \bm A_h \bm u_h = 0,
 \quad t_{s-1} < t \leq t_s,
\end{equation}   
\begin{equation}\label{12}
 \bm u_h(t_{s-1}) = \bm u_h^s ,
\end{equation} 
where $h$ is the discretization parameter. The main feature of the problems we are considering is that the matrices $\bm A_h$ and $\bm B_h$ 
are real and asymmetric.

The modal approximation corresponds to the representation of the approximate solution  ($\bm u_h \approx \bm u_N$) of problem   (\ref{11}), (\ref{12}) in the following form
\begin{equation}\label{13}
 \bm u_N(\bm x, t) =
 \sum_{n=1}^{N} a_n(t) \bm w_n(\bm x),
\end{equation} 
where $N$ is the number of dominant eigenvalues of the spectral problem, 
$\bm w_n(\bm x)$ --- corresponding eigenfunctions.

Let us define eigenfunctions and eigenvalues as the solution of the  $\alpha$-eigenvalue problem:
\begin{equation}\label{14}
 \bm A_h \bm v = \lambda  \bm B_h \bm v .
\end{equation} 
In the simplest case all eigenvalues of the spectral problem  (\ref{14}) are real:
\[
 \lambda_1 \leq \lambda_2 \leq ... \leq \lambda_{N_h} .
\] 
Under these conditions \citep{Laub2005,Ortega1987} the general solution of equation  (\ref{11}) is
\begin{equation}\label{15}
 \bm u_h(\bm x, t) =
 \sum_{n=1}^{N_h} b_n \exp(-\lambda_n (t-t_{s-1})) \bm v_n(\bm x) , 
\end{equation} 
that is in (\ref{13}) 
\[
 a_n(t) = b_n \exp(-\lambda_n (t-t_{s-1})) ,
 \quad \bm w_n(\bm x) = \bm v_n(\bm x),
 \quad n = 1,2, ..., N .  
\] 

In the general case, eigenfunctions and eigenvalues of the spectral problem  (\ref{14}) are complex.
Taking into account the reality of the matrix coefficients  $\bm A_h, \ \bm B_h$ complex eigenvalues appear as pairs of complex conjugate numbers. For example, we have a pair of $n,n+1$: 
\[
 \lambda_{n+1} = \mathrm{Re} \lambda_n - i \mathrm{Im} \lambda_n . 
\] 
Then in the representation  (\ref{13}) we obtain
\[
\begin{split}
 a_n(t) \bm w_n(\bm x) & = b_n \mathrm{Re} \big ( \exp(-\lambda_n (t-t_{s-1})) \bm v_n(\bm x) \big ), \\
 a_{n+1}(t) \bm w_{n+1}(\bm x) & = b_{n+1} \mathrm{Im} \big ( \exp(-\lambda_n (t-t_{s-1})) \bm v_n(\bm x) \big ) .
\end{split}
\] 

A special attention should be paid to define the coefficients $a_n(t_{s-1}) = b_n, \ n = 1,2, ..., N$.
For this, the initial condition (\ref{12}) is involved. For example, in the case of real eigenvalues, we have
\[
 \bm u_h^s (\bm x) = \sum_{n=1}^{N_h} b_n \bm v_n(\bm x) .
\] 
This representation is not very suitable for practical use with modal approximation, when we work only with dominant eigenfunctions.  

The initial condition includes two components  $\bm u_h^s (\bm x) = (\bm \phi_h^s (\bm x), \bm c_h^s (\bm x))$.
Dynamic behaviour of these components is due to different time-scale processes. Delayed neutrons source determines slow processes, when  $\bm c(\bm x,t)$ changes slightly with the reactor state change. In contrast, neutron flux $\bm \phi(\bm x,t)$ determines fast processes when the reactor state changes. By virtue of this separation of dynamic processes, we model the slow phase of the dynamics of the reactor with modal approximation and orientate ourselves on the approximate prediction of the initial state for delayed neutrons, only the function  $\bm c_h^s (\bm x)$ is approximated. The approximation  $\bm \phi_h^s (\bm x)$ is not of interest to us; we do not model a fast phase of the state change.

The standard approach for the decomposition of the function  $\bm u_h^s (\bm x) $ 
over the system of non-orthogonal functions $\bm v_n(\bm x), \ n = 1,2, ..., N_h$ 
consists in using the biorthogonal system of functions 
\citep{henry1975nuclear,brezinski1991biorthogonality}. 
Consider the spectral problem adjoint to (\ref{14})  
\begin{equation}\label{16}
 \bm A_h^T \widetilde{\bm v}  = \lambda  \bm B_h^T \widetilde{\bm v} .
\end{equation} 
The eigenfunctions of problems  (\ref{14}) and (\ref{16}) are orthogonal \citep{Laub2005,Ortega1987}  in the sense of the equality
\[
  (\bm B_h \bm v_n, \widetilde{\bm v}_m)= 0, 
  \quad m \neq n,
  \quad m, n = 1,2, ..., N_h , 
\] 
where $(\cdot, \cdot)$ means corresponding scalar product. 
In view of this, one can obtain
\begin{equation}\label{17}
 b_n = \frac{1}{(\bm B_h \bm v_n, \widetilde{\bm v}_n)} (\bm u_h^s, \bm B_h \widetilde{\bm v}_n),
 \quad n = 1,2, ..., N_h .  
\end{equation} 
With known solutions of the spectral problems  (\ref{14}), (\ref{16})  
the solution is represented in the form (\ref{15}), (\ref{17}).

In the approximate solution of problem   (\ref{11}), (\ref{12}) only the first $N$ coefficients  $b_n$ in (\ref{17}) are used (see (\ref{13})):
\begin{equation}\label{18}
 \bm c_h^s (\bm x) \approx  \sum_{n=1}^{N} b_n \bm c_n(\bm x) ,
\end{equation} 
where $\bm v_n (\bm x) = (\bm \phi_n (\bm x), \bm c_n (\bm x))$.
In this case, the spectral problems  (\ref{14}), (\ref{16}) are solved for $N$ dominant 
eigenvalues.

The solution of the adjoint spectral problem is involved only for calculating the initial condition coefficients. This complication of the problem is not always justified. Therefore, it is worth to use simpler algorithms for obtaining the coefficients  $b_n, \ n = 1,2, ..., N$ in (\ref{18}).
We can define them, for example, based on linear least squares  \citep{LSPk1996,verdu2014modal}.
In this case, one can obtain
\begin{equation}\label{19}
 (\bm r_N, \bm r_N) \longrightarrow \min, 
 \quad \bm r_N (\bm x)  = \bm c_h^s (\bm x) -  \sum_{n=1}^{N} b_n \bm c_n(\bm x) .
\end{equation} 
To find the coefficients, a system of linear equations is solved. 

The state change modal method is based on the following calculating scheme.
\begin{description}
 \item[Off-line calculation.] Calculation of the coefficients of the mathematical model of the multigroup diffusion approximation for the isolated reactor states, which is performed in advance. The status passport also includes calculated dominant eigenvalues and eigenfunctions of the  $\alpha$-eigenvalue problem (\ref{14}). 
These data can be supplemented by dominant eigenvalues and eigenvalues of the conjugate eigenvalue problem (\ref{16}).
 \item[On-line calculation.] Real-time modeling is carried out on the basis of the modal solution of the problem  (\ref{11}), (\ref{12}).
The coefficients in the representation  (\ref{18}) are calculated from the initial condition using (\ref{17})
or (\ref{19}). The solution for other time intervals is determined according to (\ref{15}).   
\end{description}  

\section{Computational aspects of the state change modal method} 

In practical implementation of this approach for approximate solution of reactor dynamic problems, the key aspects are spatial approximation and numerical solution of spectral problems. We will not discuss the available opportunities in this direction, but will give only a brief description of how these issues are solved for the below calculations. 

To approximate in space variables, we use the finite element method  \citep{brenner,quarteroni}. 
Let $H^1(\Omega)$ is the Sobolev space consisting of scalar functions $v$ such that  $v^2$ and  $\vert\nabla v\vert^2$ have a finite integral in  $\Omega$. For vector functions  $\bm v = \{v_1, v_2, ..., v_d\}$ we define similarly  $V^d = [H^1(\Omega)]^d$.  
For test functions we use the notation  $\bm \chi = \{\chi_1, \chi_2, ..., \chi_G\}$,
$\bm s = \{s_1, s_2, ..., s_M\}$. 
In the variational form, problem  (\ref{5}) has the following form: find $\bm \phi \in V^D, \ \bm c \in V^M$, for which
\begin{equation}\label{20}
\begin{split}
 \int_\Omega \left (V \frac{d \bm\phi}{d t} + S\bm \phi \right )\bm \chi  d\bm x 
 & + \sum_{g=1}^{G} \int_\Omega D_g\nabla \phi_g \nabla \chi_g  d\bm x 
 + \sum_{g=1}^{G} \int_{\partial \Omega} \gamma_g \phi_g \chi_g  d\bm x \\
 & = \int_\Omega R \bm \phi\bm \chi d\bm x + \int_\Omega B\bm c\bm s d\bm x, \\
 \int_\Omega \frac{d \bm{c}}{d t} \bm s d\bm x 
 &+ \int_\Omega \Lambda \bm{c} \bm s d\bm x = \int_\Omega Q \bm{\phi} \bm \chi d\bm x
\end{split}
\end{equation}
for all $\bm \chi  \in V^D, \ \bm s \in V^M$.

Further, we must pass from the continuous variational problem  (\ref{20}) to the discrete problem. We introduce finite-dimensional finite element spaces  $V_h^D \subset V^D$, $V_h^M \subset V^M$.
The discrete variational problem is formulated as follows: find  $\bm \phi^h \in V_h^D, \ \bm c^h \in V_h^M$, such that
\begin{equation}\label{21}
\begin{split}
 \int_\Omega \left (V \frac{d \bm\phi^h}{d t} + S\bm \phi^h \right )\bm \chi^h  d\bm x 
 & + \sum_{g=1}^{G} \int_\Omega D_g\nabla \phi_g^h \nabla \chi_g^h  d\bm x 
 + \sum_{g=1}^{G} \int_{\partial \Omega} \gamma_g \phi_g^h \chi_g^h  d\bm x \\
 & = \int_\Omega R \bm \phi^h \bm \chi^h d\bm x + \int_\Omega B\bm c^h \bm s^h d\bm x, \\
 \int_\Omega \frac{d \bm{c}^h}{d t} \bm s^h d\bm x 
 &+ \int_\Omega \Lambda \bm{c}^h \bm s^h d\bm x = \int_\Omega Q \bm{\phi}^h \bm \chi^h d\bm x
\end{split}
\end{equation}
for all $\bm \chi^h  \in V_h^D, \ \bm s^h \in V_h^M$.
For two-dimensional problems, scalar functions (components of vector functions) are approximated on a triangular grid using Lagrangian finite elements with polynomials of degree 1, 2 and 3. The calculation of the first eigenvalues and the corresponding eigenfunctions is a standard problem in computational mathematics \citep{Saadbook}.
It is necessary to note some fundamental features of the spectral problems (\ref{14}) и (\ref{16}).
The first feature is related to the fact that the problems under consideration are large-scale problems: two-dimensional or three-dimensional in space and many unknown quantities (a system of equations). This means that in applied problems, we have to focus on the use of computers of parallel architecture. The second is due to the asymmetry of the matrices. This leads to the appearance of complex roots.

In our study  (see \cite{avvakumov2017spectral,nd-mm}) we focus on the use of well-designed algorithms and relevant free software. To solve spectral problems with non-symmetrical matrices we use the SLEPc 
(Scalable Library for Eigenvalue Problem Computations, http://slepc.upv.es/). 
This library has traditionally been widely used (see, for instance,  \cite{hernandez2003resolution,hernandez2005slepc})
for numerical solution of the spectral problems in nuclear reactor calculations. We use a Krylov-Schur algorithm, a variation of Arnoldi method, proposed by \citep{stewart2002krylov}. 

\section{The test: the dynamics of the VVER-1000 reactor during the transition from the supercritical mode to the subcritical mode} 

A test problem for a VVER-1000 reactor without a reflector  \citep{chao} 
is considered in the two-dimensional approximation ($\Omega$ is a reactor core cross-section). 

\subsection{General description} 

\begin{figure}[!h]
  \begin{center}
    \includegraphics[width=0.75\linewidth] {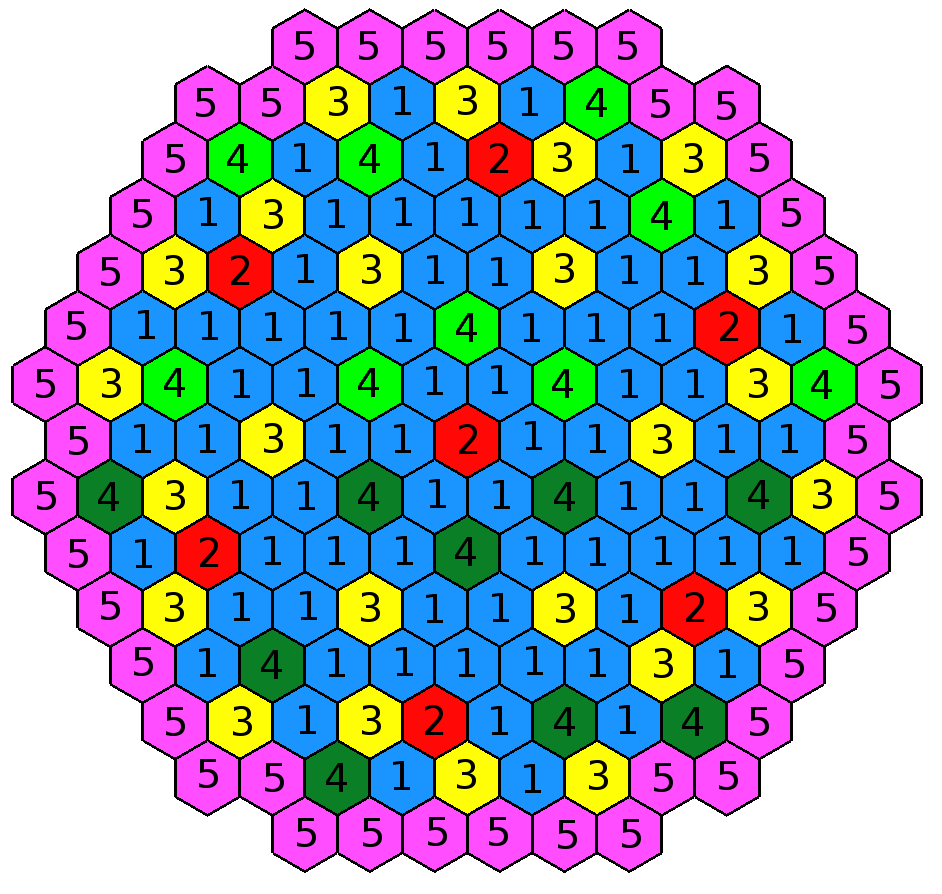}
	\caption{Geometrcial model of the VVER-1000 reactor core.}
	\label{fig:2}
  \end{center}
\end{figure}

The geometric model of the VVER-1000 core consists of a set of hexagonal-shaped cassettes and is shown in Fig.\ref{fig:2}, where fuel assemblies of various types are shown. The assembly \emph{wrench}  size is 23.6 cm.

\begin{figure}[!h]
  \begin{center}
\begin{minipage}{0.30\linewidth}
\center{\includegraphics[width=1\linewidth]{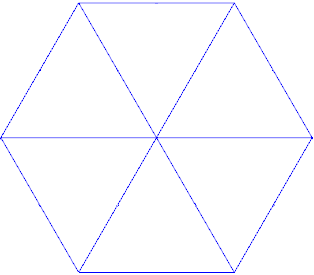}}\\
\end{minipage}
\hfill
\begin{minipage}{0.30\linewidth}
\center{\includegraphics[width=1\linewidth]{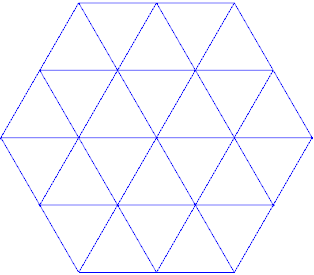}}\\
\end{minipage}
\hfill
\begin{minipage}{0.30\linewidth}
\center{\includegraphics[width=1\linewidth]{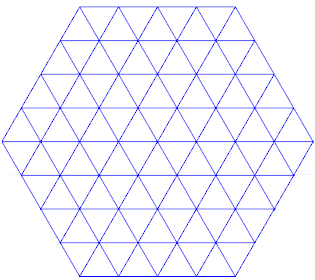}}\\
\end{minipage}
\caption{Discretization of assembly into 6, 24 and 96 finite elements.}
\label{fig:3}
  \end{center}
\end{figure}

For an approximate solution of the problem, regular triangular grids are used. The number of triangles per cassette $\kappa$  varies from 6 to 96 (Fig.\ref{fig:3}). 
A two-group approximation is used taken into account delayed neutrons: 
\begin{equation}\label{22}
\begin{split}
 \frac{1}{v_1}  \frac{\partial \varphi_1}{\partial t}  
 & - \nabla \cdot D_1 \nabla \varphi_1  + \Sigma_1 \varphi_1 + \Sigma_{s,1\rightarrow 2} \varphi_1  \\
 & - (1 - \beta_1)(\nu \Sigma_{f1} \varphi_1 + \nu \Sigma_{f2} \varphi_2) - \lambda_1 s = 0, \\
 \frac{1}{v_2}  \frac{\partial \varphi_2}{\partial t}  
 & - \nabla \cdot D_2 \nabla \varphi_2  + \Sigma_2 \varphi_2 - \Sigma_{s,1\rightarrow 2} \varphi_1   = 0,\\
 \frac{\partial s}{\partial t} & + \lambda_1 s - \beta_1(\nu \Sigma_{f1} \varphi_1 + \nu \Sigma_{f2} \varphi_2) = 0. 
\end{split}
\end{equation} 

\begin{table}[htp]
\caption{Diffusion neutronics constants for VVER-1000}
\label{t-1}
\begin{center}
\begin{tabular}{|c|c|c|c|c|c|}
\hline
Material & 1 & 2 & 3 & 4 & 5\\
\hline
$D_1$ & 1.38320e-0 & 1.38299e-0  & 1.39522e-0  & 1.39446e-0  & 1.39506e-0 \\
$D_2$ & 3.86277e-1 & 3.89403e-1 & 3.86225e-1 & 3.87723e-1 & 3.84492e-1 \\
$\Sigma_1 + \Sigma_{s,1\rightarrow 2}$ & 2.48836e-2 & 2.62865e-2 & 2.45662e-2 & 2.60117e-2 & 2.46141e-2\\
$\Sigma_2$ & 6.73049e-2 & 8.10328e-2 & 8.44801e-1 & 9.89671e-2 & 8.93878e-2\\
$\Sigma_{s,1\rightarrow 2}$ & 1.64977e-2 & 1.47315e-2 & 1.56219e-2 & 1.40185e-2 & 1.54981e-2\\
$\nu\Sigma_{f1}$ & 4.81619e-3 & 4.66953e-3 & 6.04889e-3 & 5.91507e-3 & 6.40256e-3\\
$\nu\Sigma_{f2}$ & 8.46154e-2 & 8.52264e-2 & 1.19428e-1 & 1.20497e-1 & 1.29281e-1\\
\hline
\end{tabular}
\end{center}
\end{table}

The supercritical state of the reactor is characterized by a set of coefficients, which are given in Table \ref{t-1}. 
The following boundary conditions (\ref{3}) are used:  $\gamma_g = 0.5, \ g = 1,2$.  
The following delayed neutrons parameters are used: one group of delayed neutrons with effective fraction $\beta_1 = 6.5\cdot10^{-3}$ and decay constant $\lambda_1 = 0.08$ s$^{-1}$. 
Neutron velocity  $v_1 = 1.25 \cdot 10^7$ cm/s and $v_2 = 2.5 \cdot 10^5$ cm/s.

\subsection{Supercritical state: $\alpha$-eigenvalue problem} 

Below the results of a numerical solution of the $\alpha$-eigenvalue problem (\ref{14}) are presented. 
In the framework of the two-group approximation and taking into account the delayed neutrons, one can write 
\begin{equation}\label{23}
\begin{split}
 - \nabla \cdot D_1 \nabla \varphi_1  + \Sigma_1 \varphi_1 + \Sigma_{s,1\rightarrow 2} \varphi_1  & \\
 - (1 - \beta_1)(\nu \Sigma_{f1} \varphi_1 + \nu \Sigma_{f2} \varphi_2) - \lambda_1 s & = \lambda^{(\alpha)} \frac{1}{v_1}   \varphi_1, \\
 - \nabla \cdot D_2 \nabla \varphi_2  + \Sigma_2 \varphi_2  - \Sigma_{s,1\rightarrow 2} \varphi_1  
 & = \lambda^{(\alpha)} \frac{1}{v_2}   \varphi_2,\\
 \lambda_1 s - \beta_1(\nu \Sigma_{f1} \varphi_1 + \nu \Sigma_{f2} \varphi_2) & = \lambda^{(\alpha)} s. 
\end{split}
\end{equation} 
The dominant eigenvalues $\alpha_n = \lambda_n^{(\alpha)}, \ n = 1,2, ..., N$ are searched for at
\[
 \mathrm{Re}  \lambda_1^{(\alpha)} \leq  \mathrm{Re}  \lambda_2^{(\alpha)} \leq ... 
 \leq \mathrm{Re}  \lambda_N^{(\alpha)} \leq ...\, \leq \mathrm{Re}  \lambda_{N_h}^{(\alpha)}.
\]
Similar calculations of the eigenvalues for the VVER-1000 test problem without delayed neutrons can be found in \cite{avvakumov2017spectral}. 

\begin{table}[h]
\caption{Eigenvalues $\alpha_n = \lambda_n^{(\alpha )}, \ n = 1,2, ..., 5$}
\label{t-2}
\begin{center}
\begin{tabular}{cccccc}
\hline
$p$ & $\kappa$ & $\alpha_1$ &  $\alpha_2, \alpha_3$ &  $\alpha_4, \alpha_5$ \\ 
\hline
& 6 & -0.22557  & 0.04241 $\mp$ 3.08808e-06$i$  & 0.06588 $\mp$ 4.80449e-07$i$  \\
1 & 24 & -0.82690  & 0.03777 $\mp$ 5.37884e-06$i$  & 0.06489 $\mp$ 1.37315e-06$i$ \\
& 96 & -1.74998  & 0.03619 $\mp$ 5.69002e-06$i$  & 0.06456 $\mp$ 1.40299e-06$i$ \\
\hline
& 6 & -2.10154  & 0.03592 $\mp$ 4.96474e-06$i$  & 0.06452 $\mp$ 1.21320e-06$i$ \\
2 & 24 & -2.46601  & 0.03562 $\mp$ 5.78277e-06$i$  & 0.06445 $\mp$ 1.40897e-06$i$ \\
& 96 & -2.50375  & 0.03559 $\mp$ 5.80693e-06$i$  & 0.06444 $\mp$ 1.41324e-06$i$ \\
\hline
& 6 & -2.47975  & 0.03561 $\mp$ 5.83718e-06$i$  & 0.06445 $\mp$ 1.41869e-06$i$ \\
3 & 24 & -2.50294  & 0.03559 $\mp$ 5.80783e-06$i$  & 0.06444 $\mp$ 1.41341e-06$i$ \\
& 96 & -2.51280  & 0.03558 $\mp$ 5.80954e-06$i$  & 0.06444 $\mp$ 1.41362e-06$i$ \\
\hline
\end{tabular}
\end{center}
\end{table}

\begin{table}[h]
\caption{Eigenvalues $\alpha_n = \lambda_n^{(\alpha )}, \ n = 6,7, ..., 10$}
\label{t-3}
\begin{center}
\begin{tabular}{ccccccc}
\hline
$p$ & $\kappa$ & $\alpha_6$ &  $\alpha_7$ & $\alpha_8$ &  $\alpha_9, \alpha_{10}$ \\ 
\hline
& 6 & 0.07107  & 0.07214  & 0.07323  & 0.07397 $\mp$ 2.04990e-08$i$ \\
1 & 24 & 0.07050  & 0.07167  & 0.07283  & 0.07362 $\mp$ 3.65907e-08$i$ \\
& 96 & 0.07033  & 0.07152  & 0.07269  & 0.07351 $\mp$ 3.91936e-08$i$  \\
\hline
& 6  & 0.07030  & 0.07151  & 0.07268  & 0.07349 $\mp$ 3.69824e-08$i$ \\
2 & 24 & 0.07027  & 0.07147  & 0.07265  & 0.07347 $\mp$ 4.03121e-08$i$ \\
& 96  & 0.07026  & 0.07147  & 0.07265  & 0.07347 $\mp$ 4.02324e-08$i$ \\
\hline
& 6 & 0.07027  & 0.07147  & 0.07265  & 0.07347 $\mp$ 4.02573e-08$i$ \\
3 & 24 & 0.07026  & 0.07147  & 0.07265  & 0.07347 $\mp$ 4.02248e-08$i$ \\
& 96 & 0.07026  & 0.07147  & 0.07265  & 0.07347 $\mp$ 4.02332e-08$i$ \\
\hline
\end{tabular}
\end{center}
\end{table}

The results of solving the spectral problem (\ref{23}) for the first eigenvalues  $\alpha_n = \lambda_n^{(\alpha)}, \ n = 1,2, ..., N$, $ N=10$
on different computational grids using different finite element approximations are shown in Table \ref{t-2}, \ref{t-3}. The eigenvalues $\alpha_2, \alpha_3$, $\alpha_4, \alpha_5$, $\alpha_9, \alpha_{10}$ 
of the spectral problem (\ref{23}) are complex with small imaginary parts, the eigenvalues $\alpha_1, \alpha_6$, $\alpha_7, \alpha_8$ are real.

In our example, the main eigenvalue is negative and therefore the major harmonic will increase, and all others will fade. This demonstrates the regular mode of the reactor operation. The value $\alpha = \lambda_1^{(\alpha)}$ itself determines the neutron flux amplitude and is directly related to the reactor period in the regular regime.

\begin{figure}[!h]
  \begin{center}
\begin{minipage}{0.49\linewidth}
\center{\includegraphics[width=1\linewidth]{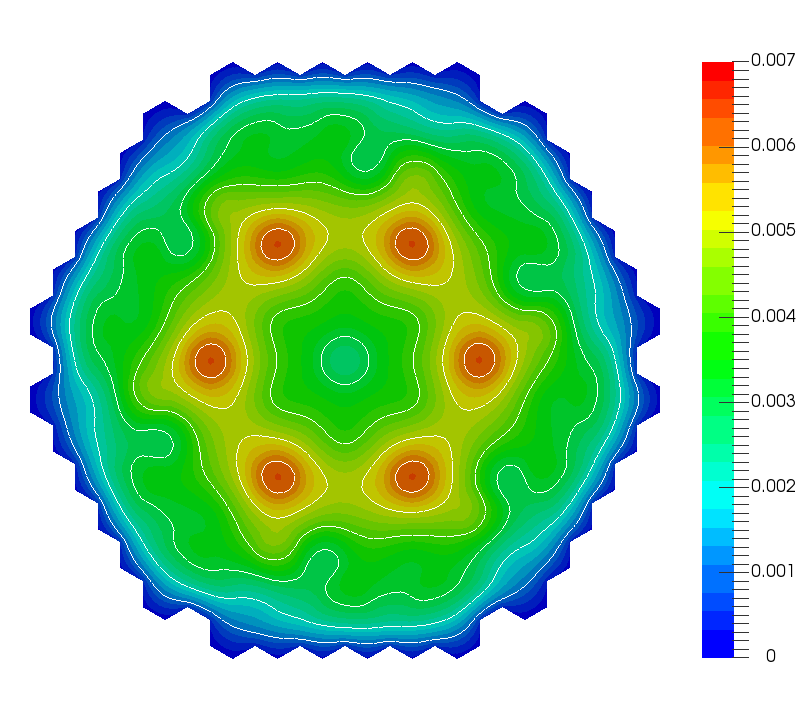}} \\
\end{minipage}
\hfill
\begin{minipage}{0.49\linewidth}
\center{\includegraphics[width=1\linewidth]{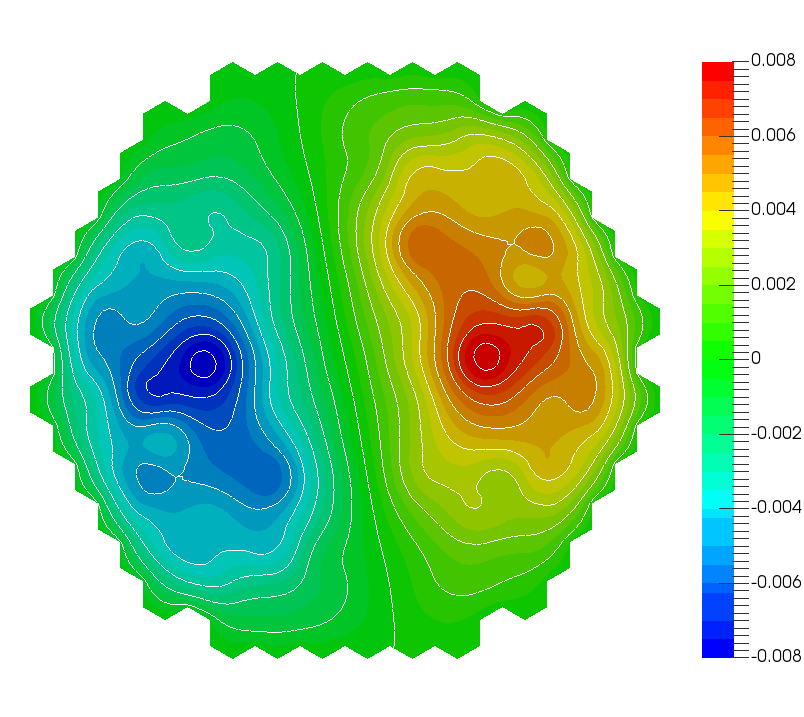}} \\
\end{minipage}
\caption{The eigenfunction $\varphi^{(1)}_1$ (left) and real part of eigenfunctions $\varphi^{(2)}_1, \ \varphi^{(3)}_1$  (right).}
\label{fig:4}
  \end{center}
\end{figure}

\begin{figure}[!h]
  \begin{center}
\begin{minipage}{0.49\linewidth}
\center{\includegraphics[width=1\linewidth]{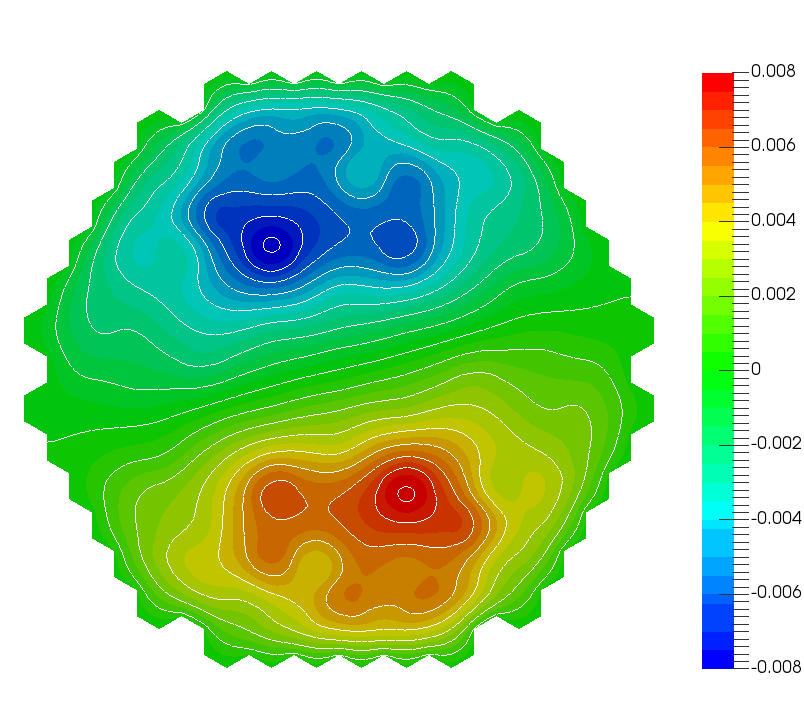}} \\
\end{minipage}
\hfill
\begin{minipage}{0.49\linewidth}
\center{\includegraphics[width=1\linewidth]{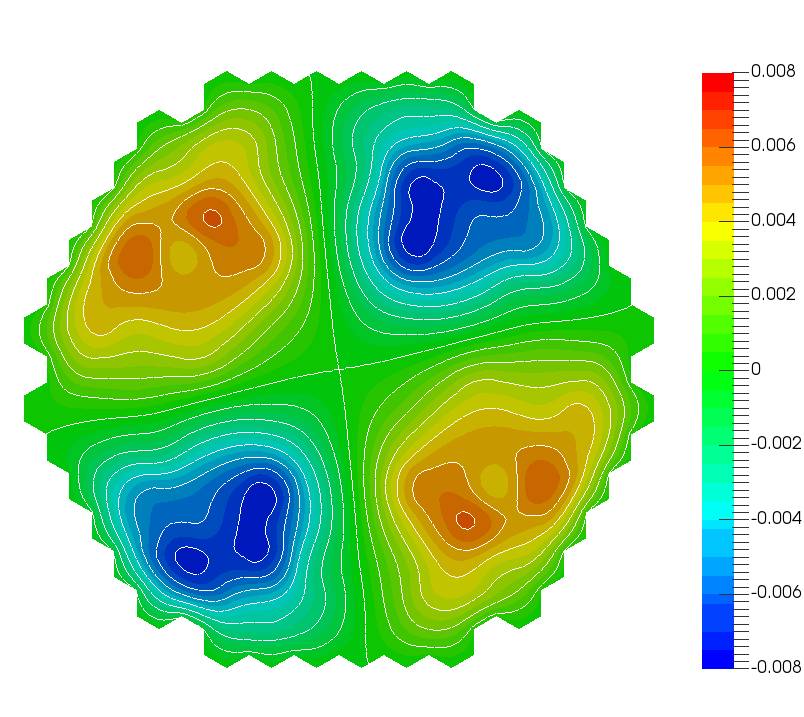}} \\
\end{minipage}
\caption{Imaginary part of eigenfunctions $\varphi^{(2)}_1, \ - \varphi^{(3)}_1$ (left) and  real part of eigenfunctions $\varphi^{(4)}_1, \ \varphi^{(5)}_1$  (right).}
\label{fig:5}
  \end{center}
\end{figure}

\begin{figure}[!h]
  \begin{center}
\begin{minipage}{0.49\linewidth}
\center{\includegraphics[width=1\linewidth]{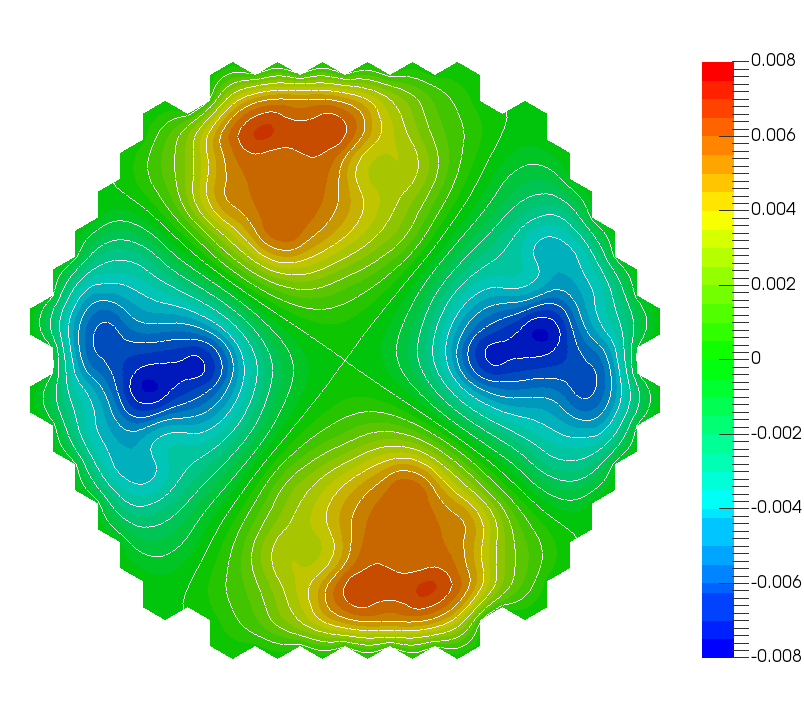}} \\
\end{minipage}
\hfill
\begin{minipage}{0.49\linewidth}
\center{\includegraphics[width=1\linewidth]{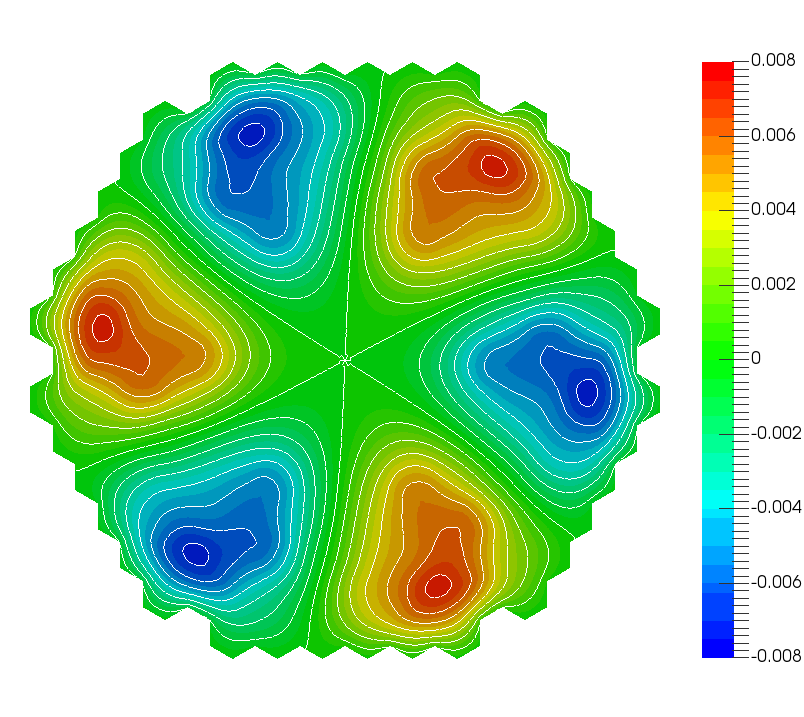}} \\
\end{minipage}
\caption{Imaginary part of eigenfunctions $\varphi^{(4)}_1, \ - \varphi^{(5)}_1$ (left) and  eigenfunction $\varphi^{(6)}_1$  (right).}
\label{fig:6}
  \end{center}
\end{figure}

\begin{figure}[!h]
  \begin{center}
\begin{minipage}{0.49\linewidth}
\center{\includegraphics[width=1\linewidth]{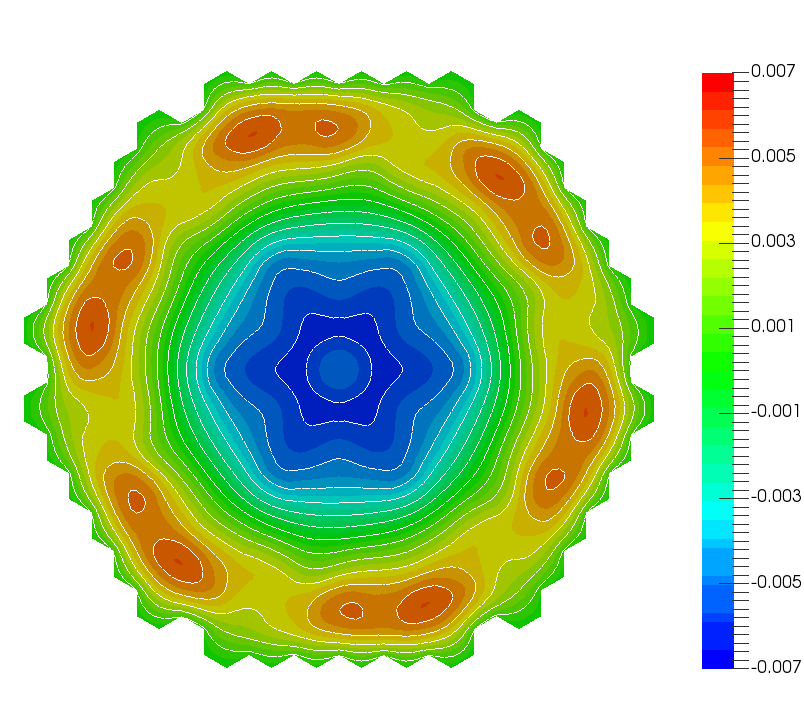}} \\
\end{minipage}
\hfill
\begin{minipage}{0.49\linewidth}
\center{\includegraphics[width=1\linewidth]{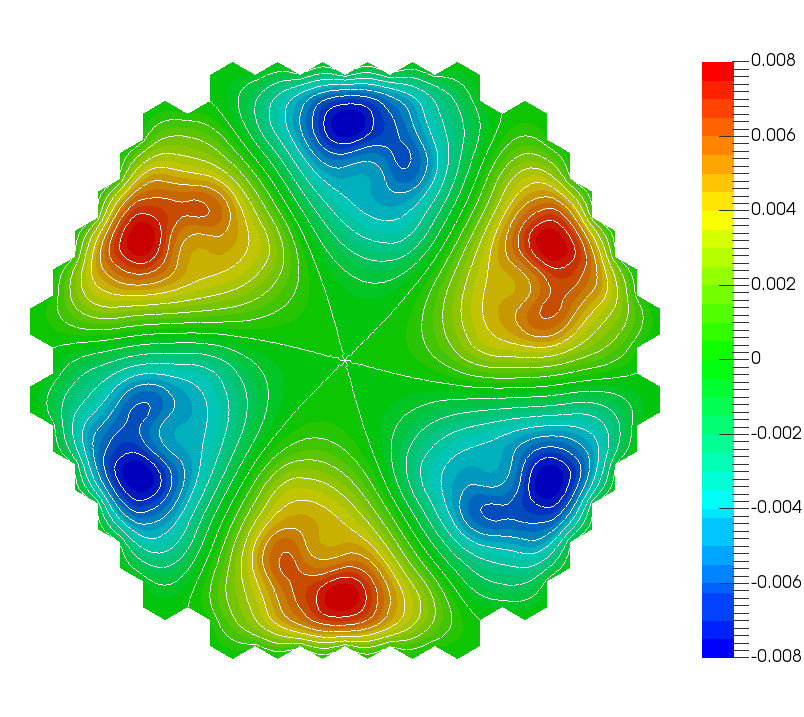}} \\
\end{minipage}
\caption{The eigenfunction $\varphi^{(7)}_1$ (left) and  eigenfunction $\varphi^{(8)}_1$  (right).}
\label{fig:7}
  \end{center}
\end{figure}

\begin{figure}[!h]
  \begin{center}
\begin{minipage}{0.49\linewidth}
\center{\includegraphics[width=1\linewidth]{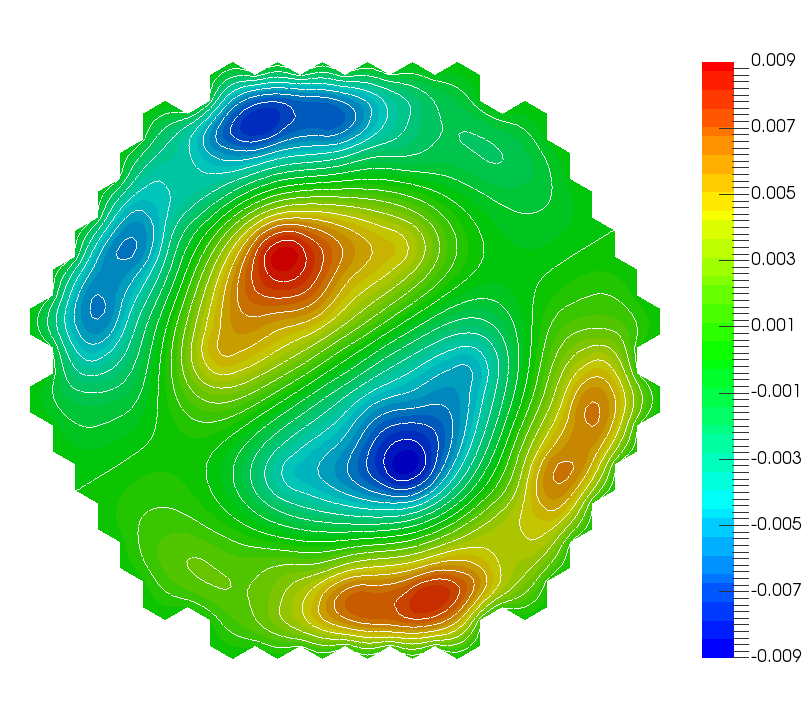}} \\
\end{minipage}
\hfill
\begin{minipage}{0.49\linewidth}
\center{\includegraphics[width=1\linewidth]{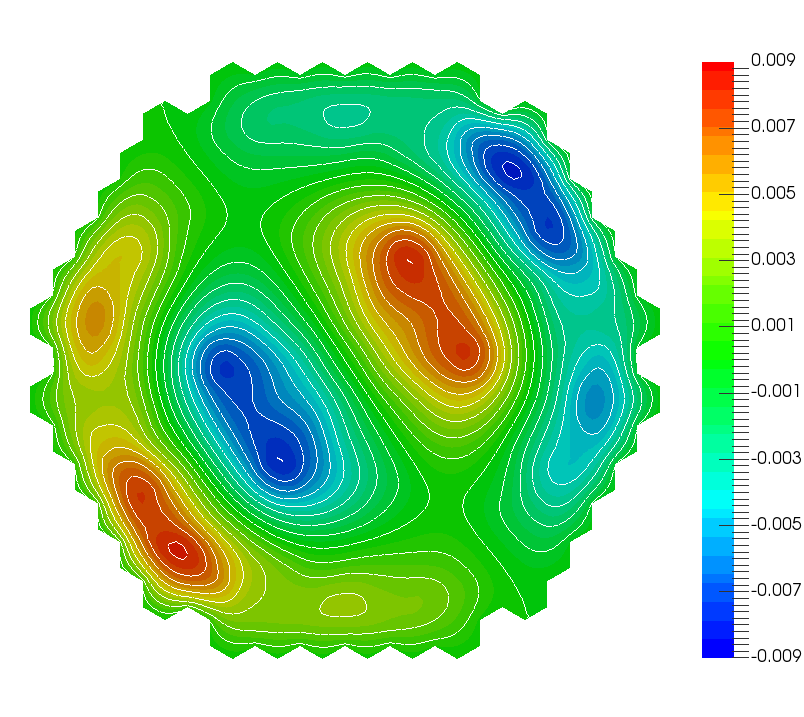}} \\
\end{minipage}
\caption{Real part of eigenfunctions $\varphi^{(9)}_1, \ \varphi^{(10)}_1$ (left) and  imaginary part of eigenfunctions $\varphi^{(9)}_1, \ - \varphi^{(10)}_1$  (right).}
\label{fig:8}
  \end{center}
\end{figure}

\begin{figure}[!h]
  \begin{center}
\begin{minipage}{0.49\linewidth}
\center{\includegraphics[width=1\linewidth]{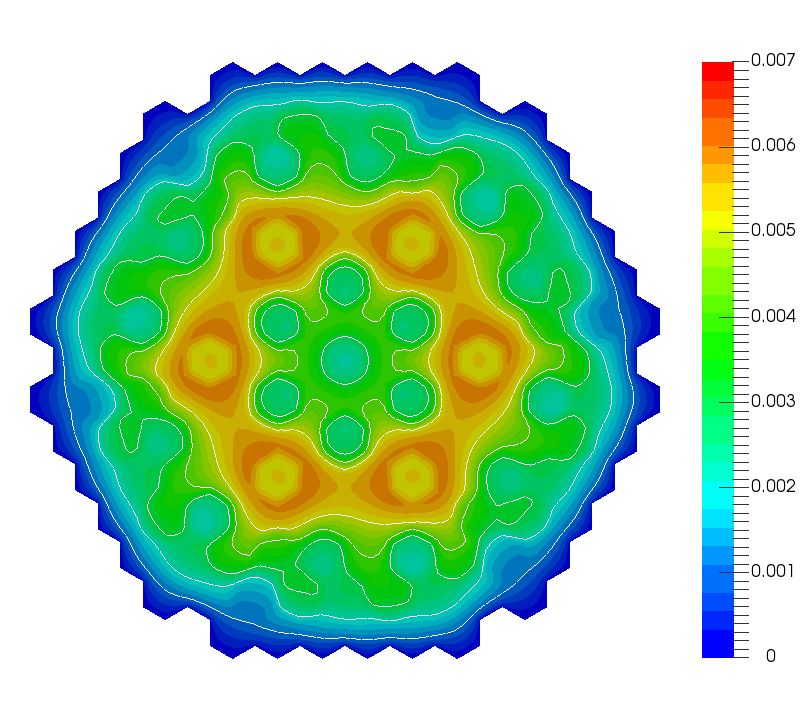}} \\
\end{minipage}
\hfill
\begin{minipage}{0.49\linewidth}
\center{\includegraphics[width=1\linewidth]{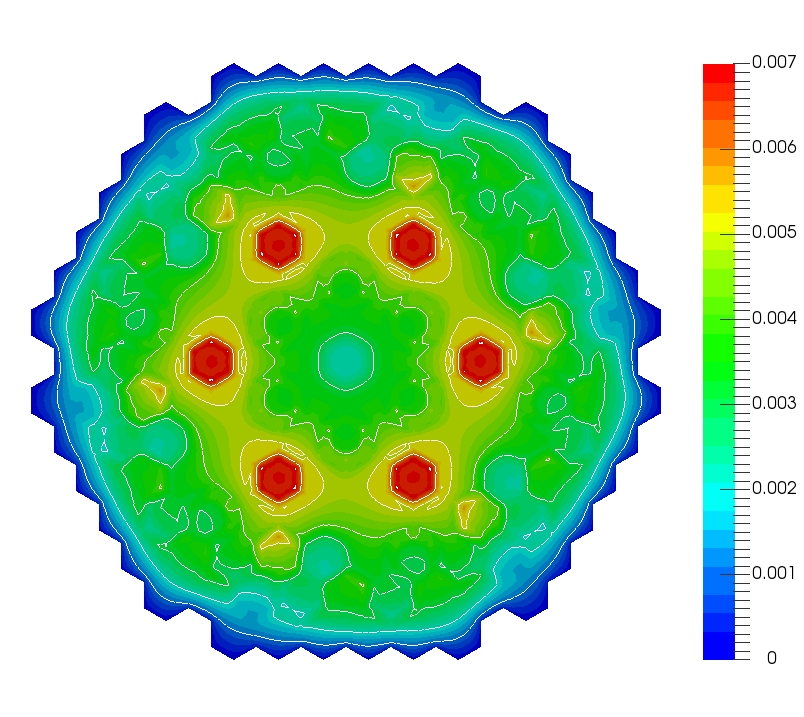}} \\
\end{minipage}
\caption{The eigenfunction $\varphi^{(1)}_2$ (left) and the eigenfunctions $s^{(1)}$  (right).}
\label{fig:9}
  \end{center}
\end{figure}

Dominant eigenfunctions of the spectral problem (\ref{23}) for the first group are shown in Figs.\ref{fig:4}--\ref{fig:8}. 
The calculations are performed on a grid with $\kappa=96$ using finite elements of degree $p=3$.
The main eigenfunctions for the second group of  $\varphi^{(1)}_2$ and delayed neutrons source $s^{(1)}$  are shown in Fig.\ref{fig:9}. 

\begin{table}[h]
\caption{Eigenvalues $\alpha_n = \lambda_n^{(\alpha )}, \ n = 1,2, ..., 10$
for direct and adjoint problems}
\label{t-4}
\begin{center}
\begin{tabular}{rll}
\hline
$n$ & $\alpha_n$ for problem (\ref{14}) & $\alpha_n$ for problem (\ref{16}) \\
\hline
1 & -2.51280117966 & -2.51280117972 \\
2,3 & 0.0355815000364 $\mp$ 5.80954455861e-06 & 0.0355815000365 $\mp$ 5.80954421646e-06 \\
4,5 & 0.0644427013767 $\mp$ 1.41362187449e-06 & 0.0644427013767 $\mp$ 1.41362190730e-06 \\
6 & 0.0702618501639 & 0.0702618501639 \\
7 & 0.0714652882224 & 0.0714652882164 \\
8 & 0.0726456060606 & 0.0726456060606 \\
9,10 & 0.0734708921578 $\mp$ 4.02332269037e-08 & 0.0734708921578 $\mp$ 4.02332146248e-08 \\
\hline
\end{tabular}
\end{center}
\end{table}

\subsection{Adjoint spectral problem} 

Analogous data were obtained for approximate solution of the adjoint spectral problem (\ref{16}).
The eigenvalues of the spectral problems (\ref{14}) and (\ref{16}) coincide. Their difference from each other is an indirect measure of the accuracy of the numerical solution. Data on the dominant eigenvalues, which are given in Table \ref{t-4} ($k=96, \ p = 3$), show that the eigenvalues of the main and adjoint spectral problems are close to each other with good accuracy.

The spectral problems under consideration are characterized by small imaginary parts of the eigenvalues. Therefore, we can expect that the eigenfunctions of problem (\ref{14}) 
are close to orthogonal. As illustration, Table \ref{t-5} contains data for the scalar products $(\phi_1^{(n)}, \phi_1^{(m)})$
for the first 10 eigenfunctions. For convenience of comparison, the eigenfunctions are normalized in $L_2(\Omega)$:
\[
 \phi_1^{(n)} \longrightarrow \frac{\phi_1^{(n)}}{\|\phi_1^{(n)}\|} .
\] 
The maximum non-orthogonality (for $(\phi_1^{(1)}, \phi_1^{(7)})$)
does not exceed 10 \%.
The biorthogonality condition of the eigenfunctions of the fundamental functions (see (\ref{14})) and the adjoint spectral problems (see (\ref{16})) is valid with approximately the same accuracy.
This observed error can be related to an approximate calculation of eigenvalues and eigenfunctions.

\begin{table}[h]
\caption{Scalar product $(\phi_1^{(n)}, \phi_1^{(m)}), \ n, m = 1,2, ..., 10$}
\label{t-5}
\begin{center}
\footnotesize 
%\small
\begin{tabular}{c|rrrrrrrrrr}
\hline
$n$\textbackslash$m$&1&2&3&4&5&6&7&8&9&10 \\
\hline
1 & 1.0e-00 & 1.3e-08 & 2.2e-08 & -3.8e-08 & 9.8e-09 & -1.8e-09 & 1.0e-02 & -3.2e-09 & -2.2e-08 & 1.6e-09 \\ 
2 & 1.3e-08 & 1.0e-00 & -1.6e-08 & -1.6e-08 & 1.4e-08 & 4.1e-08 & 1.2e-09 & -2.0e-07 & -3.1e-03 & 7.5e-03 \\ 
3 & 2.2e-08 & -1.6e-08 & 1.0e-00 & -9.8e-09 & -1.1e-08 & -1.8e-08 & 1.1e-08 & -3.3e-08 & -7.5e-03 & -3.1e-03 \\ 
4 & -3.8e-08 & -1.6e-08 & -9.8e-09 & 1.0e-00 & -3.9e-10 & -1.1e-08 & 1.4e-08 & 4.0e-09 & 3.0e-09 & -1.1e-08 \\ 
5 & 9.8e-09 & 1.4e-08 & -1.1e-08 & -3.9e-10 & 1.0e-00 & 2.9e-09 & -1.6e-08 & -1.9e-08 & 6.3e-09 & 6.3e-09 \\ 
6 & -1.8e-09 & 4.1e-08 & -1.8e-08 & -1.1e-08 & 2.9e-09 & 1.0e-00 & -4.2e-09 & -5.6e-03 & 4.1e-08 & -1.2e-07 \\ 
7 & 1.0e-02 & 1.2e-09 & 1.1e-08 & 1.4e-08 & -1.6e-08 & -4.2e-09 & 1.0e-00 & -2.1e-09 & -1.8e-08 & 8.0e-09 \\ 
8 & -3.2e-09 & -2.0e-07 & -3.3e-08 & 4.0e-09 & -1.9e-08 & -5.6e-03 & -2.1e-09 & 1.0e-00 & -5.2e-08 & 2.3e-07 \\ 
9 & -2.2e-08 & -3.1e-03 & -7.5e-03 & 3.0e-09 & 6.3e-09 & 4.1e-08 & -1.8e-08 & -5.2e-08 & 1.0e-00 & -5.5e-07 \\ 
10 & 1.6e-09 & 7.5e-03 & -3.1e-03 & -1.1e-08 & 6.3e-09 & -1.2e-07 & 8.0e-09 & 2.3e-07 & -5.5e-07 & 1.0e-00 \\ 
\hline
\end{tabular}
\end{center}
\end{table}

Within the modal method, we can not rely on high accuracy when considering a relatively small number of dominant eigenvalues. Therefore, in the example under consideration, we can assume that the eigenvalues are real, and the corresponding eigenfunctions are orthogonal. Instead of (\ref{17}) coefficients are used 
\begin{equation}\label{24}
 b_n \approx  \frac{1}{(\bm c_n, \bm c_n)} (\bm c_h^s, \bm c_n),
 \quad n = 1,2, ..., N ,
\end{equation}
to approximate the initial condition.

\subsection{Subcritical state} 

In the supercritical mode, due to the sufficiently large magnitude of the main eigenvalue, the regular regime of the reactor is rapidly developing, where 
\[
 \bm u (\bm x, t) \approx a_1 \exp(-\alpha_1 t) \bm v_1^0 (\bm x) .
\] 
Here  $\bm v_1^0 (\bm x)$ is the first mode of the supercritical state. We consider the problem with the transition from this supercritical state at  $t_0 = 0$  to the subcritical state.

The subcritical stage is characterized by a 15\% increase in the coefficient  
$\Sigma_2$ for material 4 in the VVER-1000 test diffusion constants (see Table  \ref{t-1}). 
Thus, the dynamics of the reactor is as follows: 
\[
 \Sigma_2 \longrightarrow 1.15 \Sigma_2 \quad (\mathrm{material} \ 4).
\] 
The initial state is characterized by specifying the initial conditions at $t_0 = 0$ as
\begin{equation}\label{25}
 \bm u (\bm x, 0) = \bm v_1^0 (\bm x) . 
\end{equation} 

The calculational results of  the dominant eigenvalues for the subcritical state are presented in Tables \ref{t-6}, \ref{t-7}. In this case even the first eigenvalues not significantly differ from each other. 

\begin{table}[h]
\caption{Subcritical state: $\alpha_n = \lambda_n^{(\alpha )}, \ n = 1,2, ..., 5$}
\label{t-6}
\begin{center}
\begin{tabular}{cccccc}
\hline
$p$ & $\kappa$ & $\alpha_1$ &  $\alpha_2, \alpha_3$ &  $\alpha_4, \alpha_5$ \\ 
\hline
   & 6 & 0.03602 & 0.05760 $\mp$ 1.49652e-06$i$ & 0.06890 $\mp$ 4.92606e-07$i$ \\ 
1 & 24 & 0.02656 & 0.05502 $\mp$ 2.06007e-06$i$ & 0.06804 $\mp$ 1.01253e-06$i$ \\ 
  & 96 & 0.02276 & 0.05411 $\mp$ 2.16813e-06$i$ & 0.06774 $\mp$ 1.03843e-06$i$ \\ 
\hline
   & 6 & 0.02250 & 0.05404 $\mp$ 1.81823e-06$i$ & 0.06772 $\mp$ 8.73562e-07$i$ \\ 
2 & 24 & 0.02144 & 0.05380 $\mp$ 2.19400e-06$i$ & 0.06765 $\mp$ 1.04253e-06$i$ \\ 
  & 96 & 0.02125 & 0.05376 $\mp$ 2.20812e-06$i$ & 0.06763 $\mp$ 1.04715e-06$i$ \\ 
\hline
   & 6 & 0.02139 & 0.05379 $\mp$ 2.22579e-06$i$ & 0.06764 $\mp$ 1.05369e-06$i$ \\ 
3 & 24 & 0.02124 & 0.05376 $\mp$ 2.20883e-06$i$ & 0.06763 $\mp$ 1.04736e-06$i$ \\ 
  & 96 & 0.02122 & 0.05376 $\mp$ 2.20951e-06$i$ & 0.06763 $\mp$ 1.04756e-06$i$ \\ 
\hline
\end{tabular}
\end{center}
\end{table}

\begin{table}[h]
\caption{Subcritical state:  $\alpha_n = \lambda_n^{(\alpha )}, \ n = 6,7, ..., 10$}
\label{t-7}
\begin{center}
\begin{tabular}{ccccccc}
\hline
$p$ & $\kappa$ & $\alpha_6$ &  $\alpha_7$ & $\alpha_8$ &  $\alpha_9, \alpha_{10}$ \\ 
\hline
   & 6 & 0.07276 & 0.07363 & 0.07369 & 0.07466 $\mp$ 2.47162e-08$i$ \\ 
1 & 24 & 0.07222 & 0.07316 & 0.07329 & 0.07429 $\mp$ 1.08814e-08$i$ \\ 
  & 96 & 0.07204 & 0.07301 & 0.07316 & 0.07417 $\mp$ 1.38093e-08$i$ \\
\hline
   & 6 & 0.07203 & 0.07300 & 0.07315 & 0.07416 $\mp$ 1.26708e-08$i$ \\ 
2 & 24 & 0.07199 & 0.07296 & 0.07312 & 0.07413 $\mp$ 1.50527e-08$i$ \\ 
  & 96 & 0.07198 & 0.07295 & 0.07312 & 0.07413 $\mp$ 1.49196e-08$i$ \\ 
\hline
   & 6 & 0.07198 & 0.07295 & 0.07312 & 0.07413 $\mp$ 1.52256e-08$i$ \\ 
3 & 24 & 0.07198 & 0.07295 & 0.07312 & 0.07413 $\mp$ 1.49141e-08$i$ \\ 
  & 96 & 0.07198 & 0.07295 & 0.07311 & 0.07413 $\mp$ 1.49178e-08$i$ \\ 
\hline
\end{tabular}
\end{center}
\end{table}

\begin{figure}[!h]
  \begin{center}
    \includegraphics[width=0.95\linewidth] {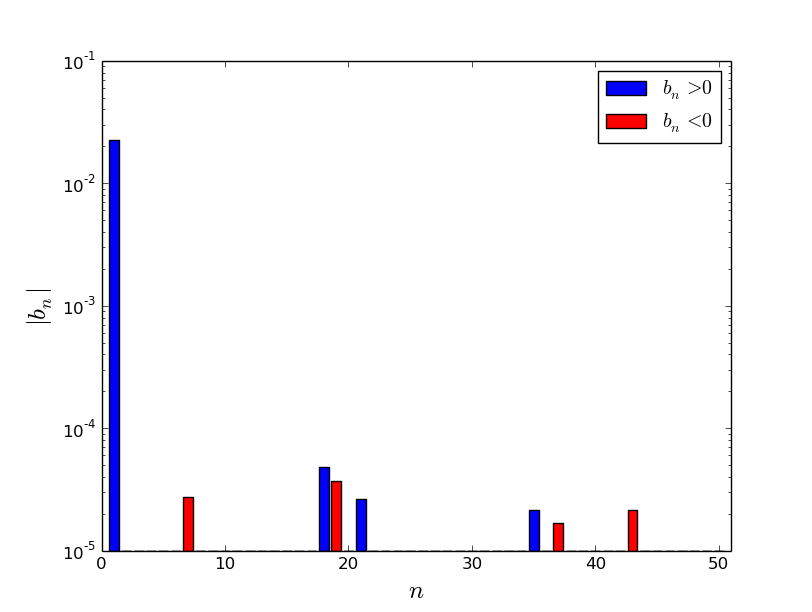}
	\caption{Approximate solution coefficients (\ref{26}).}
	\label{fig:10}
  \end{center}
\end{figure}

For an approximate solution, we use the following formulation:
\begin{equation}\label{26}
 \bm u_N(\bm x, t) = 
 \sum_{n=1}^{N} b_n \exp(- \mathrm{Re} \, \alpha_n \, t) \bm v_n(\bm x) ,  
\end{equation} 
where the coefficients $b_n, \ n = 1,2, ..., N$ are calculated according to the given initial condition (\ref{24}). These coefficients for $N=50$ are shown in Fig.\ref{fig:10}. 
As we see, an approximate solution can be described by first mode only.

We distinguish two phases of the dynamic process: fast and slow. In the fast phase, the initial condition (\ref{25}) 
is rearranged to the initial condition, which corresponds to (\ref{26}): from the function
$\bm u(\bm x, 0)$ to the function $\bm u_N(\bm x, 0)$. The slow phase is associated with the evolution of the solution according to (\ref{26}).
Within the state change modal technology, the fast phase is not modeled at all.

\begin{figure}[!h]
  \begin{center}
\begin{minipage}{0.051\linewidth}
\center{1} \\
\end{minipage}
\hfill
\begin{minipage}{0.3\linewidth}
\center{\includegraphics[width=1\linewidth]{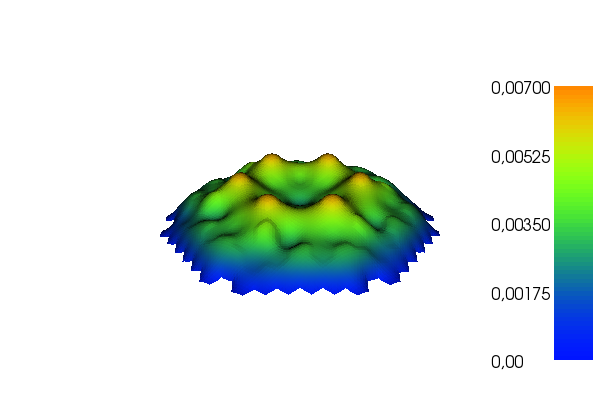}} \\
\end{minipage}
\hfill
\begin{minipage}{0.3\linewidth}
\center{\includegraphics[width=1\linewidth]{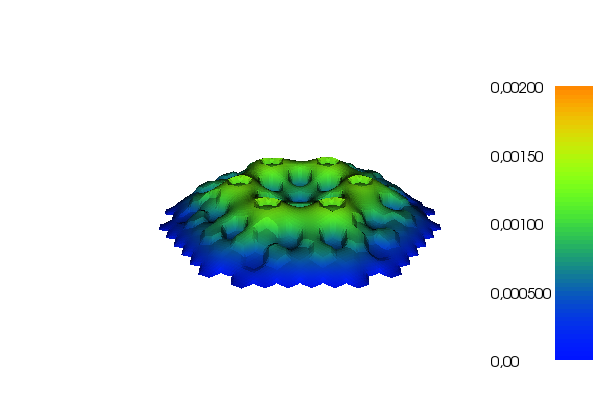}} \\
\end{minipage}
\hfill
\begin{minipage}{0.3\linewidth}
\center{\includegraphics[width=1\linewidth]{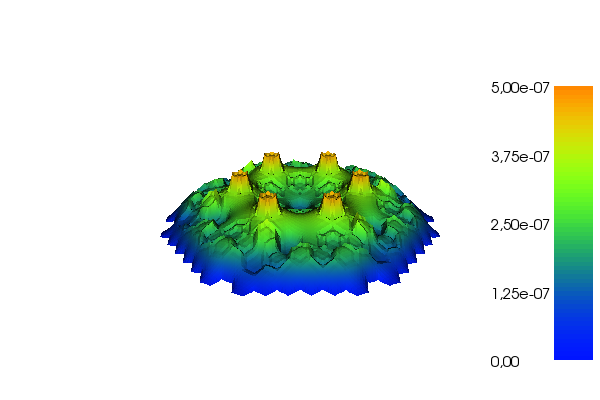}} \\
\end{minipage}

\begin{minipage}{0.051\linewidth}
\center{2} \\
\end{minipage}
\hfill
\begin{minipage}{0.3\linewidth}
\center{\includegraphics[width=1\linewidth]{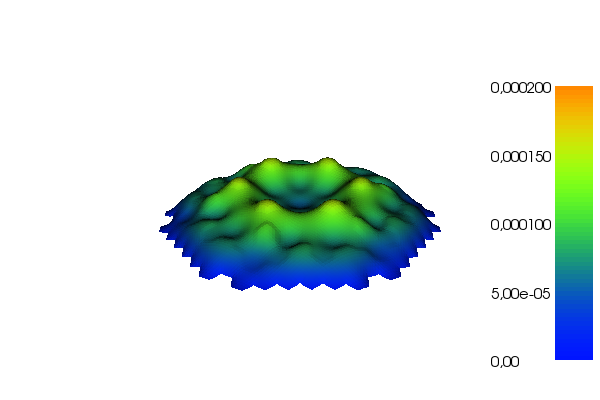}} \\
\end{minipage}
\hfill
\begin{minipage}{0.3\linewidth}
\center{\includegraphics[width=1\linewidth]{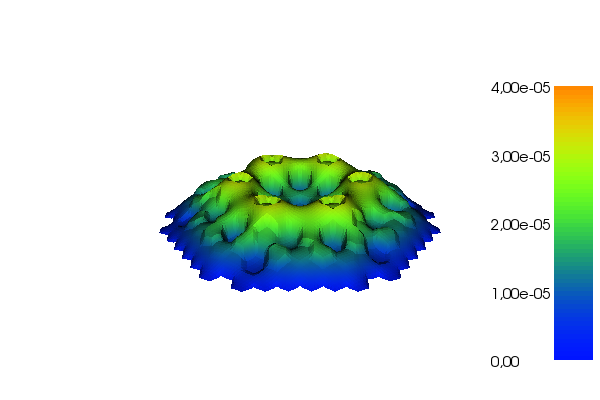}} \\
\end{minipage}
\hfill
\begin{minipage}{0.3\linewidth}
\center{\includegraphics[width=1\linewidth]{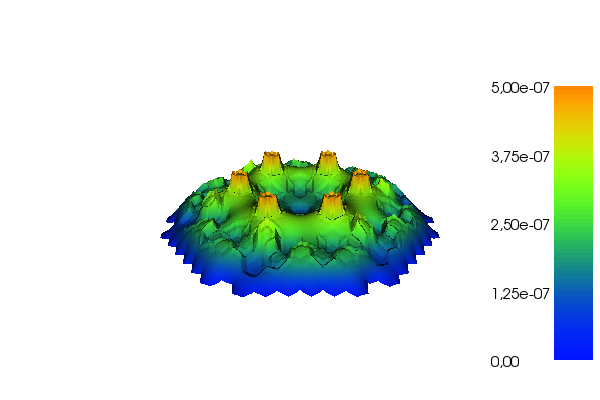}} \\
\end{minipage}

\begin{minipage}{0.051\linewidth}
\center{~} \\
\end{minipage}
\hfill
\begin{minipage}{0.3\linewidth}
\center{a} \\
\end{minipage}
\hfill
\begin{minipage}{0.3\linewidth}
\center{b} \\
\end{minipage}
\hfill
\begin{minipage}{0.3\linewidth}
\center{c} \\
\end{minipage}
\hfill

\caption{Function $\bm u(\bm x, 0)$ (line 1) and function  $\bm u_N(\bm x, 0)$ (line 2):
a --- neutron flux of group 1, b --- neutron flux of group 2, c --- delayed neutrons source.}
\label{fig:11}
  \end{center}
\end{figure}

The beginning and the end of the fast phase are illustrated through the calculational data shown in Fig. \ref{fig:11}. The results were obtained with
 $N=10$. We can note very small changes in the topology of the initial and reconstructed initial conditions. Let us pay attention to the substantial restructuring of the solution, which is illustrated by large changes in the neutron flux amplitudes for the first and second groups.

\subsection{The asymmetric perturbation} 

Consider a more complex transition to a subcritical state. The subcritical stage will be characterized by a different increase in the coefficient 
$\Sigma_2$ for material 4 in the diffusion constants in the upper and lower half of the reactor cross-section (see Fig.\ref{fig:2}). Now let the reactor dynamics corresponds to the following transformation
\[
 \Sigma_2 \longrightarrow 
 \begin{cases}
 1.1 \Sigma_2, & \mathrm{material \ 4 \ (top \ part)}, \\
 1.2 \Sigma_2, & \mathrm{material \ 4 \ (bottom \ part)}.
 \end{cases}
\] 

The calculational results for the dominant eigenvalues of the reactor subcritical asymmetric state are presented in Tables \ref{t-8}, \ref{t-9}. All eigenvalues for this reactor state are real. 

\begin{table}[h]
\caption{Subcritical asymmetric state: $\alpha_n = \lambda_n^{(\alpha )}, \ n = 1,2, ..., 5$}
\label{t-8}
\begin{center}
\begin{tabular}{ccccccc}
\hline
$p$ & $\kappa$ & $\alpha_1$ &  $\alpha_2$ & $\alpha_3$ &  $\alpha_4$ & $\alpha_5$ \\ 
\hline
   & 6 & 0.03347 & 0.05728 & 0.05788 & 0.06884 & 0.06889 \\ 
1 & 24 & 0.02333 & 0.05467 & 0.05528 & 0.06797 & 0.06802 \\ 
  & 96 & 0.01925 & 0.05374 & 0.05436 & 0.06768 & 0.06772 \\ 
\hline
   & 6 & 0.01894 & 0.05367 & 0.05429 & 0.06765 & 0.06770 \\ 
2 & 24 & 0.01782 & 0.05343 & 0.05405 & 0.06758 & 0.06762 \\ 
  & 96 & 0.01763 & 0.05339 & 0.05401 & 0.06757 & 0.06761 \\ 
\hline
   & 6 & 0.01777 & 0.05342 & 0.05404 & 0.06758 & 0.06762 \\ 
3 & 24 & 0.01762 & 0.05339 & 0.05400 & 0.06757 & 0.06761 \\ 
  & 96 & 0.01760 & 0.05338 & 0.05400 & 0.06757 & 0.06761 \\ 
\hline
\end{tabular}
\end{center}
\end{table}

\begin{table}[h]
\caption{Subcritical asymmetric state:  $\alpha_n = \lambda_n^{(\alpha )}, \ n = 6,7, ..., 10$}
\label{t-9}
\begin{center}
\begin{tabular}{cccccccc}
\hline
$p$ & $\kappa$ & $\alpha_6$ &  $\alpha_7$ & $\alpha_8$ &  $\alpha_9$ & $\alpha_{10}$ \\  
\hline
   & 6 & 0.07274 & 0.07355 & 0.07369 & 0.07464 & 0.07468 \\ 
1 & 24 & 0.07220 & 0.07309 & 0.07329 & 0.07427 & 0.07430 \\
  & 96 & 0.07202 & 0.07294 & 0.07316 & 0.07415 & 0.07419 \\ 
\hline
   & 6 & 0.07201 & 0.07293 & 0.07314 & 0.07414 & 0.07417 \\
2 & 24 & 0.07197 & 0.07289 & 0.07311 & 0.07412 & 0.07415 \\ 
  & 96 & 0.07196 & 0.07288 & 0.07311 & 0.07411 & 0.07414 \\ 
\hline
   & 6 & 0.07196 & 0.07289 & 0.07311 & 0.07411 & 0.07415 \\ 
3 & 24 & 0.07196 & 0.07288 & 0.07311 & 0.07411 & 0.07414 \\ 
  & 96 & 0.07196 & 0.07288 & 0.07311 & 0.07411 & 0.07414 \\ 
\hline
\end{tabular}
\end{center}
\end{table}

\begin{figure}[!h]
  \begin{center}
    \includegraphics[width=0.95\linewidth] {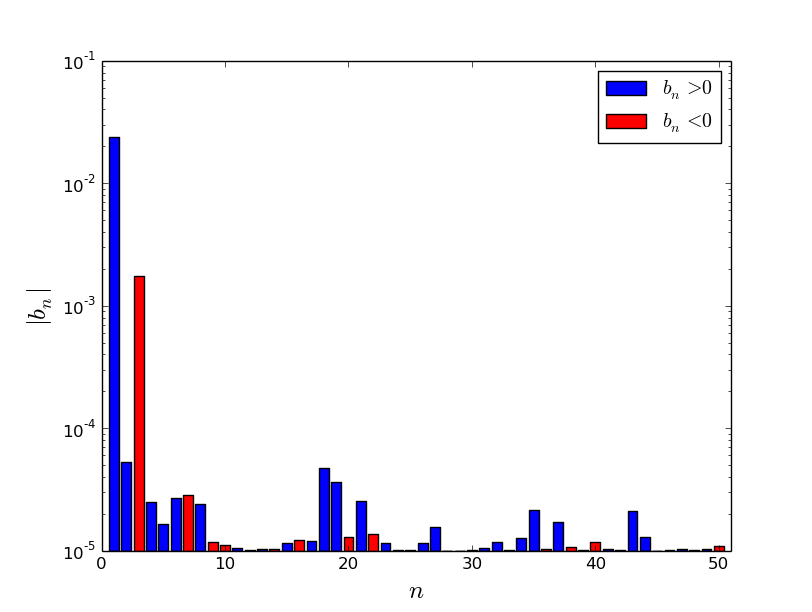}
	\caption{Approximate solution coefficients (\ref{26}) --- asymmetric perturbation.}
	\label{fig:12}
  \end{center}
\end{figure}

\begin{figure}[!h]
  \begin{center}
\begin{minipage}{0.051\linewidth}
\center{1} \\
\end{minipage}
\hfill
\begin{minipage}{0.3\linewidth}
\center{\includegraphics[width=1\linewidth]{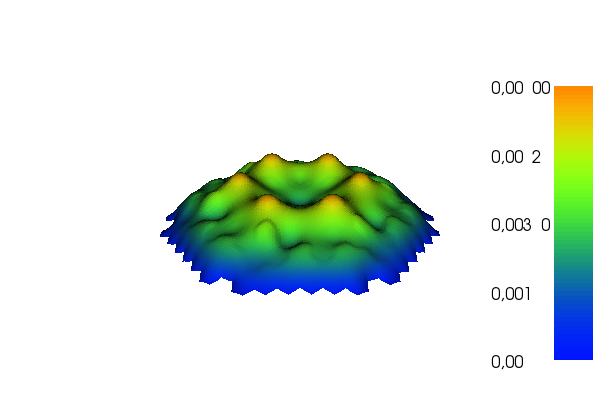}} \\
\end{minipage}
\hfill
\begin{minipage}{0.3\linewidth}
\center{\includegraphics[width=1\linewidth]{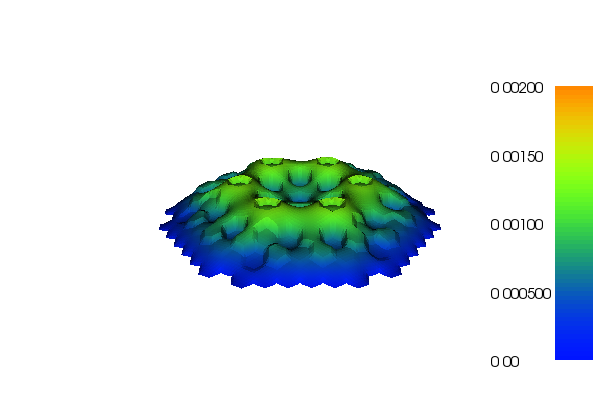}} \\
\end{minipage}
\hfill
\begin{minipage}{0.3\linewidth}
\center{\includegraphics[width=1\linewidth]{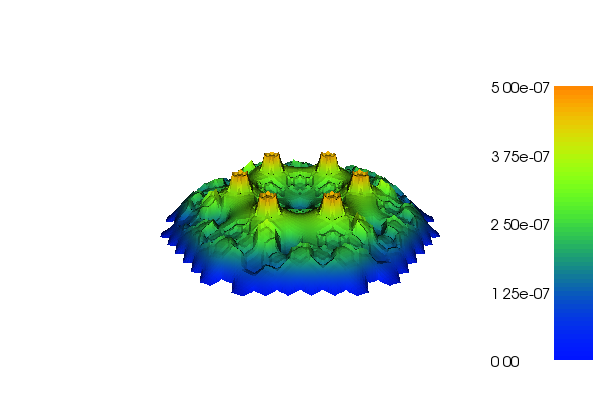}} \\
\end{minipage}

\begin{minipage}{0.051\linewidth}
\center{2} \\
\end{minipage}
\hfill
\begin{minipage}{0.3\linewidth}
\center{\includegraphics[width=1\linewidth]{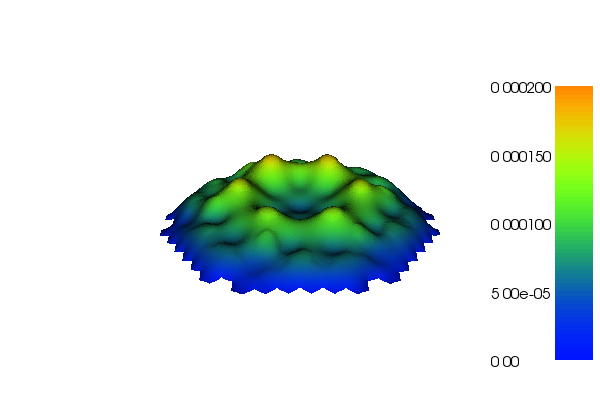}} \\
\end{minipage}
\hfill
\begin{minipage}{0.3\linewidth}
\center{\includegraphics[width=1\linewidth]{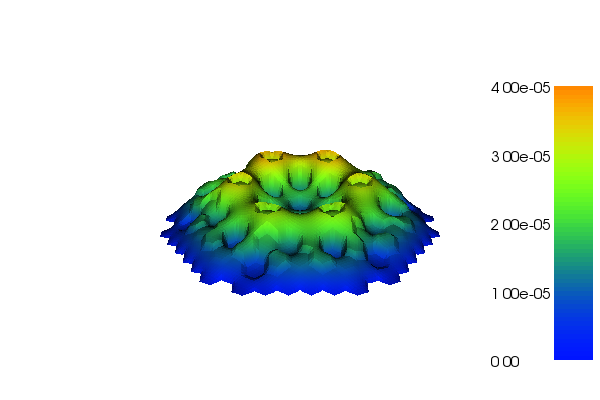}} \\
\end{minipage}
\hfill
\begin{minipage}{0.3\linewidth}
\center{\includegraphics[width=1\linewidth]{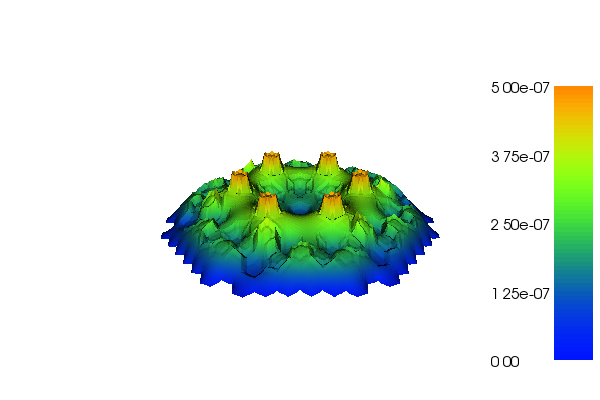}} \\
\end{minipage}

\begin{minipage}{0.051\linewidth}
\center{~} \\
\end{minipage}
\hfill
\begin{minipage}{0.3\linewidth}
\center{a} \\
\end{minipage}
\hfill
\begin{minipage}{0.3\linewidth}
\center{b} \\
\end{minipage}
\hfill
\begin{minipage}{0.3\linewidth}
\center{c} \\
\end{minipage}
\hfill

\caption{Function $\bm u(\bm x, 0)$ (line 1) and function  $\bm u_N(\bm x, 0)$ (line 2) for asymmetric perturbation:
a --- neutron flux of group 1, b --- neutron flux of group 2, c --- delayed neutrons source.}
\label{fig:13}
  \end{center}
\end{figure}

The coefficients $b_n, \ n = 1,2, ..., N$, $N=50$ of the approximate solution (\ref{26}) with the initial condition (\ref{25}) are shown in Fig.\ref{fig:12}. 
In the case under consideration, the approximate solution contains several modes and can not be described only by the first mode. The fast transition phase is illustrated in Fig. \ref{fig:13}.
Calculations of the end of the fast phase (the functions $\bm u_N(\bm x, 0)$) are performed at  
$N=10$. 

\subsection{Comparison with the nonstationary problem solution} 

\begin{figure}[!h]
  \begin{center}
    \includegraphics[width=0.75\linewidth] {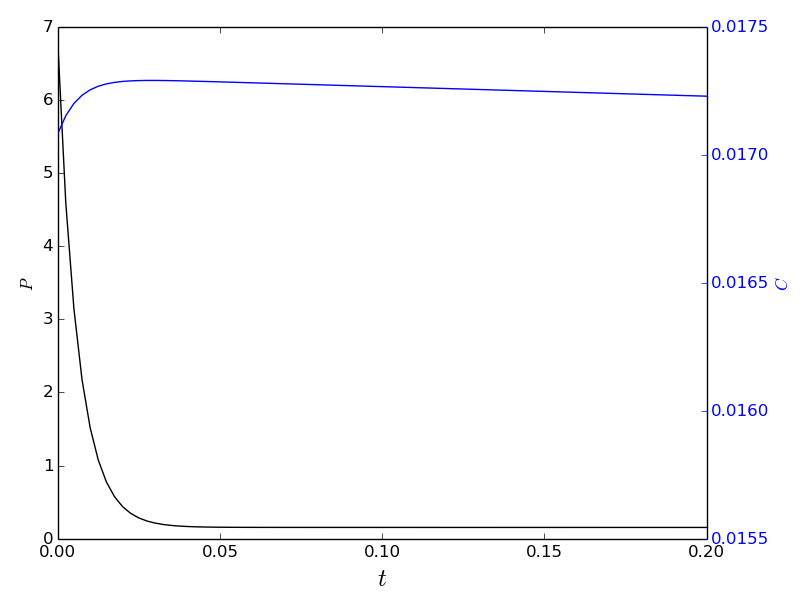}
	\caption{Fast stage of reactor state: neutronic power.}
	\label{fig:14}
  \end{center}
\end{figure}

\begin{figure}[!h]
  \begin{center}
    \includegraphics[width=0.9\linewidth] {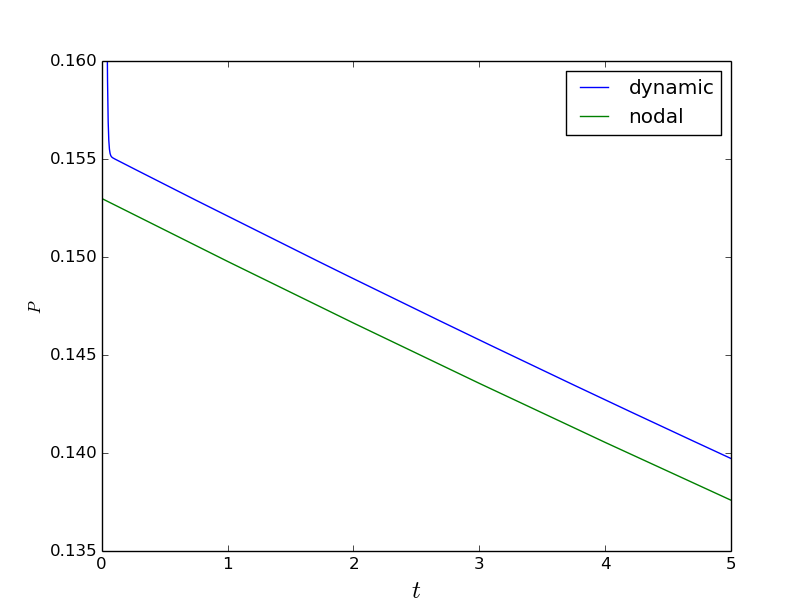}
	\caption{Slow stage of reactor state: neutronic power.}
	\label{fig:15}
  \end{center}
\end{figure}

\begin{figure}[!h]
  \begin{center}
    \includegraphics[width=0.9\linewidth] {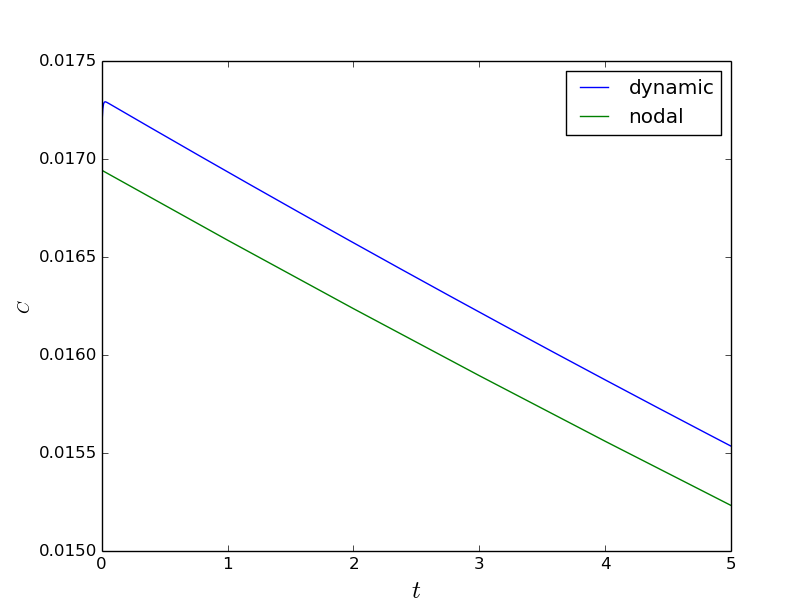}
	\caption{Slow stage of reactor state: delayed neutrons sourse.}
	\label{fig:16}
  \end{center}
\end{figure}

\begin{figure}[!h]
  \begin{center}
    \includegraphics[width=0.9\linewidth] {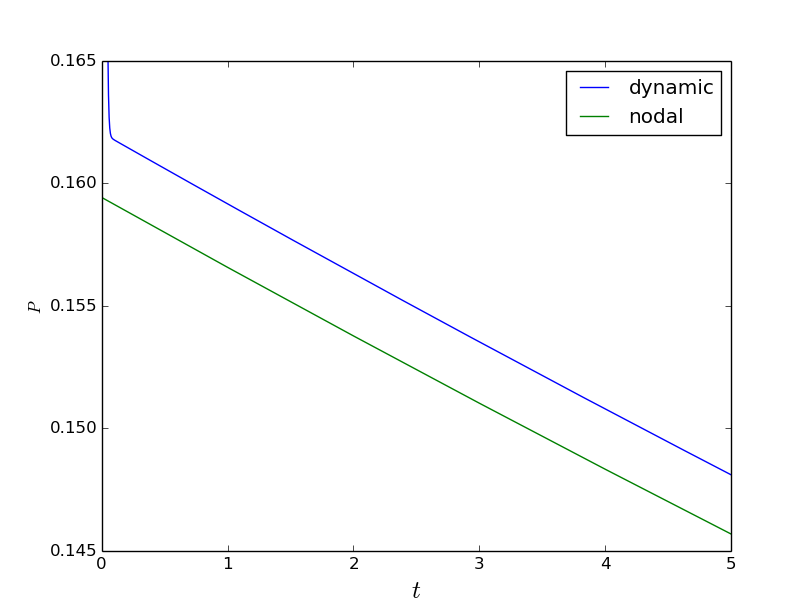}
	\caption{Slow stage of reactor state for the asymmetric perturbation: neutronic power.}
	\label{fig:17}
  \end{center}
\end{figure}

\begin{figure}[!h]
  \begin{center}
    \includegraphics[width=0.9\linewidth] {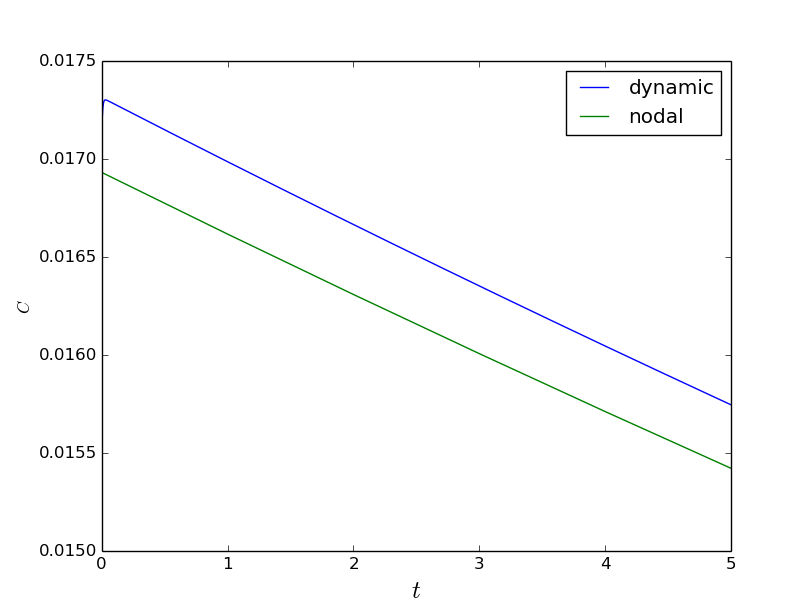}
	\caption{Slow stage of reactor state for the asymmetric perturbation: delayed neutrons source.}
	\label{fig:18}
  \end{center}
\end{figure}

An approximate solution, obtained using modal approximation, can be compared with the dynamic problem solution. The boundary value problem for the system of equations (\ref{22}) is solved. 
Fully implicit scheme on a uniform grid in time with a sufficiently small step $\tau = 0.0025$ is used (see details in
\cite{nd-mm}).
The dynamics of the neutron power of the nuclear reactor $P$ and the delayed neutrons source $C$ 
at the initial stage during the transition from the critical state to the subcritical in the case of a symmetric perturbation is shown in Fig.\ref{fig:14}. 
Here 
\[
 P(t) = \int_{\Omega} (\nu\Sigma_{f1} \varphi_1 + \nu\Sigma_{f2} \varphi_2)  d \bm x,
 \quad C(t) = \int_{\Omega} c(\bm x,t) d \bm x.
\] 
There is a rapid change in neutron power over a short period of time, while the delayed neutrons source changes rather slow. The dynamics of the slow phase is illustrated in Figs.\ref{fig:15},\ref{fig:16}. 
Here the solution of full equations (dynamic in Figs.\ref{fig:15},\ref{fig:16}),
and the modal approximation solution (modal) are presented. Similar data for the asymmetric perturbation of the reactor state are shown in Figs.\ref{fig:17},\ref{fig:18}. 
We see that the integral characteristics of the reactor dynamics at slow stage are calculated with good accuracy.

\clearpage

\section{Conclusions} 

The problem of simulation of reactor dynamic processes is considered on the basis of multigroup neutron diffusion equations accounting for delayed neutrons.
The modal approximation is used: an approximate solution is represented as an expansion on limited number of dominant eigenfunctions of the $\alpha$-eigenvalue spectral problem.

Numerical simulation of reactor non-stationary processes is carried out on the basis of a successive change in the states of the reactor. These states are characterized by a set of constant parameters to describe the multigroup neutron flux behavior.
The state change modal method was developed. The phase, which described fast transition to the approximate solution, is selected as a set of dominant modes. At a slow phase of the reactor dynamics, the solution is based on the evolution of dominant modes.

The computational implementation of the state change modal method is based on the previously calculated (of-line calculation) eigenfunctions and eigenvalues of the  $\alpha$-eigenvalue spectral problem. Fast determination of dominant modes and calculation of the reactor neutron flux at selected times are based on on-line calculation.
The classical Lagrange finite elements $p=1,2,3$ are used for the spatial approximation. 
Accuracy control is performed using condensed grids. Spectral problems are solved numerically using well-developed free software SLEPc.

Test calculations are made in two-dimensional analysis using a two-group diffusion approximation. Calculations of dominant modes for a reactor supercritical state are performed. The major mode solution, which determines the reactor regular regime, is used as the initial condition for transition to the subcritical state. The modeling of the reactor state change as a transfer from one state to another state is carried out for two cases. The first of them (symmetric perturbation) is due to the uniform change in the absorbing material properties. The second case (asymmetric perturbation) deals with a non-uniform change in the absorbing material properties (in two halves over the reactor cross-section).

Comparison of the calculational results obtained by using two methods (one based on modal approximation and another based on the full dynamics calculation) shows the acceptable accuracy in calculation of neutron power and delayed neutrons source for the VVER-1000 test problem.

\section*{Acknowledgements}

This work are supported by the Russian Foundation for Basic Research (\#~16-08-01215) 
and by the grant of the Russian Federation Government \\ (\#~14.Y26.31.0013).

\end{document}